\documentclass[showpacs, pra,twocolumn,preprintnumbers ,amsmath, amssymb, superscriptaddress, aps]{revtex4-2}
\usepackage{color}
\usepackage{amsmath,amssymb}
\usepackage{pifont}
\usepackage{amssymb}  
\usepackage{bbold}
\usepackage{float}
\usepackage{subfloat}
\usepackage[colorlinks=true,linkcolor=blue,citecolor=blue]{hyperref}%
\usepackage[caption=false]{subfig}
\usepackage{tikz}
\usepackage{makecell}
\usepackage{subfig}
\usepackage{pifont}   
\usepackage{graphicx} 
\graphicspath{{Figures/}}
\usepackage{dcolumn}  
\usepackage{bm}       
\usepackage{multirow} 
\usepackage{placeins}
\usepackage{mathtools}

\begin{document}	
	\title{Tunneling conductance through gapped bilayer graphene junctions}
	\date{\today}
	\author{Nadia Benlakhouy}
	\email{benlakhouy.n@ucd.ac.ma}
	\affiliation{Laboratory of Theoretical Physics, Faculty of Sciences, Choua\"ib Doukkali University, PO Box 20, 24000 El Jadida, Morocco}
	\author{Ahmed Jellal}
	\email{a.jellal@ucd.ac.ma}
	\affiliation{Laboratory of Theoretical Physics, Faculty of Sciences, Choua\"ib Doukkali University, PO Box 20, 24000 El Jadida, Morocco}
	\affiliation{Canadian Quantum  Research Center,
				204-3002 32 Ave Vernon,  BC V1T 2L7,  Canada}
			
				\author{El Houssine Atmani}
			\affiliation{Laboratory of Nanostructures and Advanced Materials, Mechanics and Thermofluids, Faculty of Sciences and Techniques, Hassan II University, Mohammedia, Morocco}

	\begin{abstract}
		The conductance through single-layer graphene (SLG) and AA/AB-stacked bilayer graphene (BLG) junctions is obtained by taking into account band gap and bias voltage terms. First, we consider gapped SLG, while in between, they are connected into pristine BLG. For Fermi energy larger than the interlayer hopping, the conductance as a function of the bilayer region  length $d$ reveals two different models of anti-resonances with the same period. As a function of the band gap, with AA-BLG stacking, the results show that the conductance has the same minima whatever the value of $d$, and for AB-BLG, $d$ remains relevant such that the system creates a global energy gap. Second, we consider pristine SLG, and in between, they are connected to gapped-biased BLG. We observe the appearance of peaks in the conductance profile with different periods and shapes, and also the presence of Klein tunneling with zero conductance in contrast to the first configuration. When $ d $ is less than 10, $G(E)$ vanishes and exhibits anti-Klein tunneling as a function of the Fermi energy $E$. We also investigate the conductance as a function of the bias. For AA-BLG, the results show antiresonances and diminish for a large value of the bias, independently of the bilayer region of length. In contrast, the conductance in AB-BLG has distinct characteristics in that it begins conducting with maxima for small $E$ and with minima for large $E$.

	\end{abstract}

	\maketitle
\section{Introduction}
After being experimentally isolated, single-layer graphene (SLG) has attracted constantly expanding interest due to its unique characteristics and possible applications \cite{novoselov2004electric, neto2009electronic, nair2008fine}. Recently, bilayer graphene (BLG) has also been expanded as the next attractive two-dimensional (2D) carbon material \cite{mccann2006landau, guinea2006electronic, roy1998study, castro2007biased}. As a result of its excellent electronic structure and transport properties \cite{van2013four, benlakhouy2021transport}. BLG has two different stacking arrangements: AB-BLG and AA-BLG. AB-BLG is the most stable stack that is generally examined theoretically and experimentally \cite{ohta2006controlling, goerbig2011electronic}. The AA-BLG is just a double copy of SLG, having a linear gapless energy spectrum that has piqued the attention of many theoretical researchers \cite{rakhmanov2012instabilities, chiu2013critical}. 

Recent studies have shown that graphene can create bilayer graphene flakes that are linked to single-layer graphene regions or those with nanoribbon contacts, such as  SLG/BLG/SLG interfaces \cite{gonzalez2010electronic, sahu2010effects, zambrano2017photon, yan2016spatially, clark2014energy}. Other studies have examined multiple domain walls that distinguish between numbers of layers \cite{giannazzo2012electronic, rameshti2015supercurrent} or even various stacking types \cite{pelc2015topologically, mirzakhani2018edge}. For example, J. W. González   {\it et al.} \cite{gonzalez2010electronic} evaluate the transport properties of bilayer graphene flakes using armchair nanoribbon contacts. The conductance exhibits oscillations between zero and the maximum value of the conductance. Other theoretical and experimental research on the transport properties, edge state properties, and appearance of Landau levels in such devices were the main topics \cite{puls2009interface, koshino2010interface, yin2013mono, hasegawa2012electric, hu2012edge}. Different quantum interference effects in the conductance, such as Fabry-Pérot resonances as well as Fano anti-resonances, have been observed in these systems \cite{zambrano2017photon}. 

In most of these current theoretical studies, the charge carriers behave as transitions among systems with different transport properties. As experimentally observed, at normal incidence, Klein tunneling in SLG yields a probability of $100\%$. AB-BLG exhibits anti-Klein tunneling using the two-band approximation, and this is due to the conservation of pseudospin \cite{katsnelson2006chiral, stander2009evidence}. Hence, it is important to examine the tunneling properties of these combined systems in addition to how  the transport channels affect the transport properties. We recently \cite{benlakhouy2021transport} investigated the transport properties of a rectangular potential barrier in gapped  and biased  AB-BLG. We noticed that the conductance is affected, and  the anti-Klein tunneling is no longer maintained. Using dielectric materials such as silicon carbide (SiC) \cite{rozhkov2016electronic} or hexagonal boron nitride (h-BN) \cite{zhai2016proposal}, as well as applying an external electric field \cite{zhang2009direct}, it is possible to experimentally realize the band gap $\Delta $.

\begin{figure}
	\centering
	\subfloat[]{\includegraphics[width=0.40\linewidth]{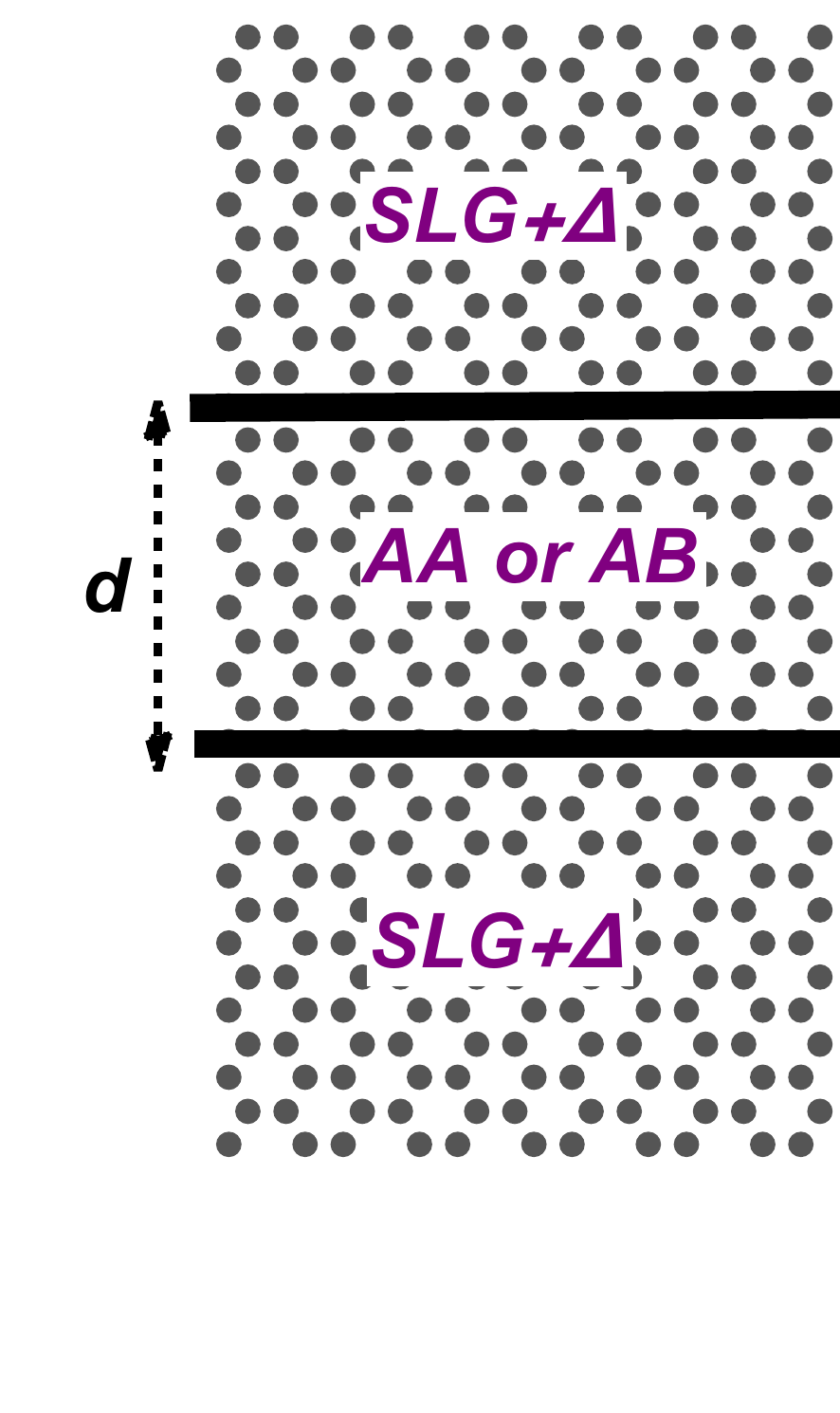}}
		\subfloat[]{\includegraphics[width=0.40\linewidth]{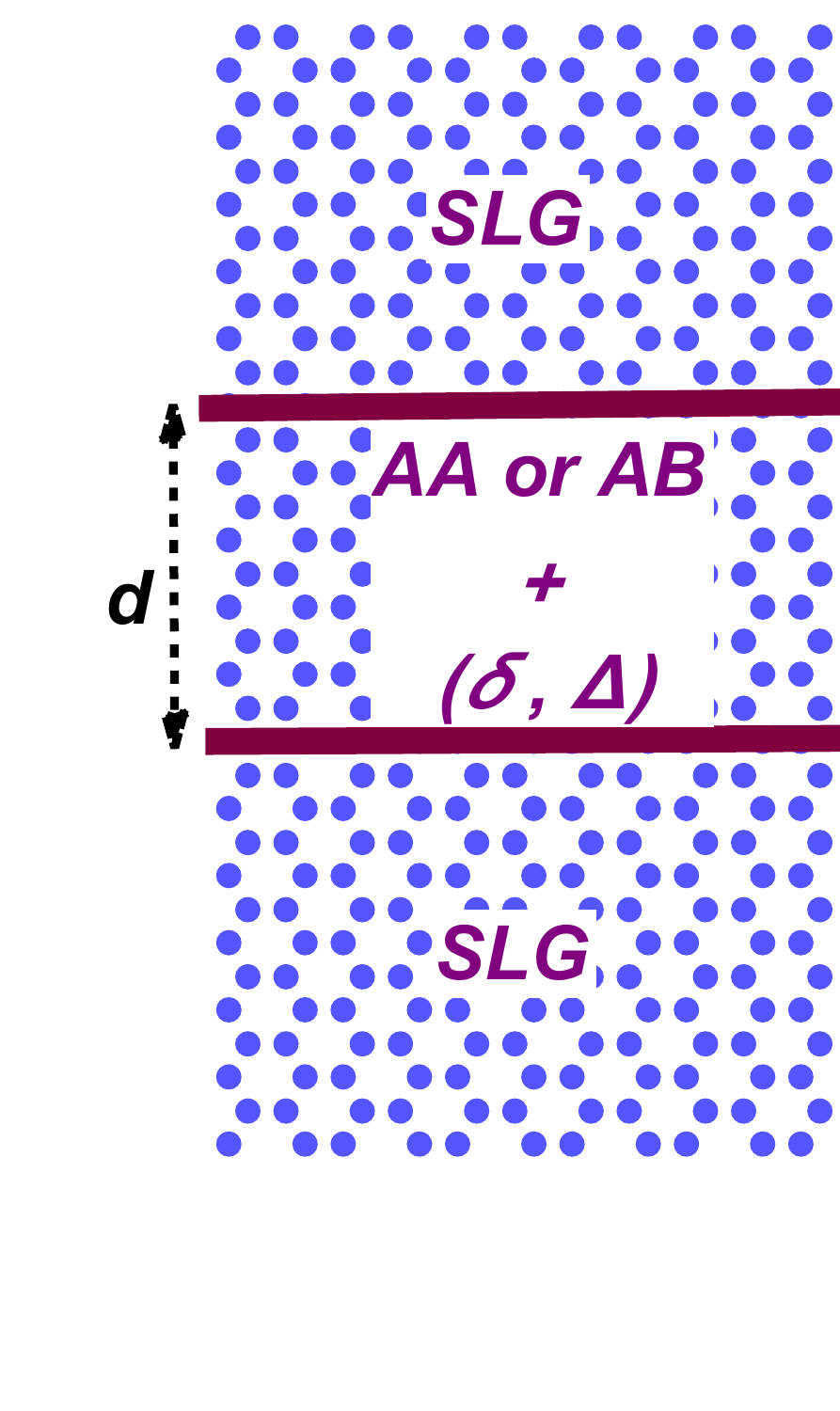}}\\ 	\subfloat[]{\includegraphics[width=0.42\linewidth]{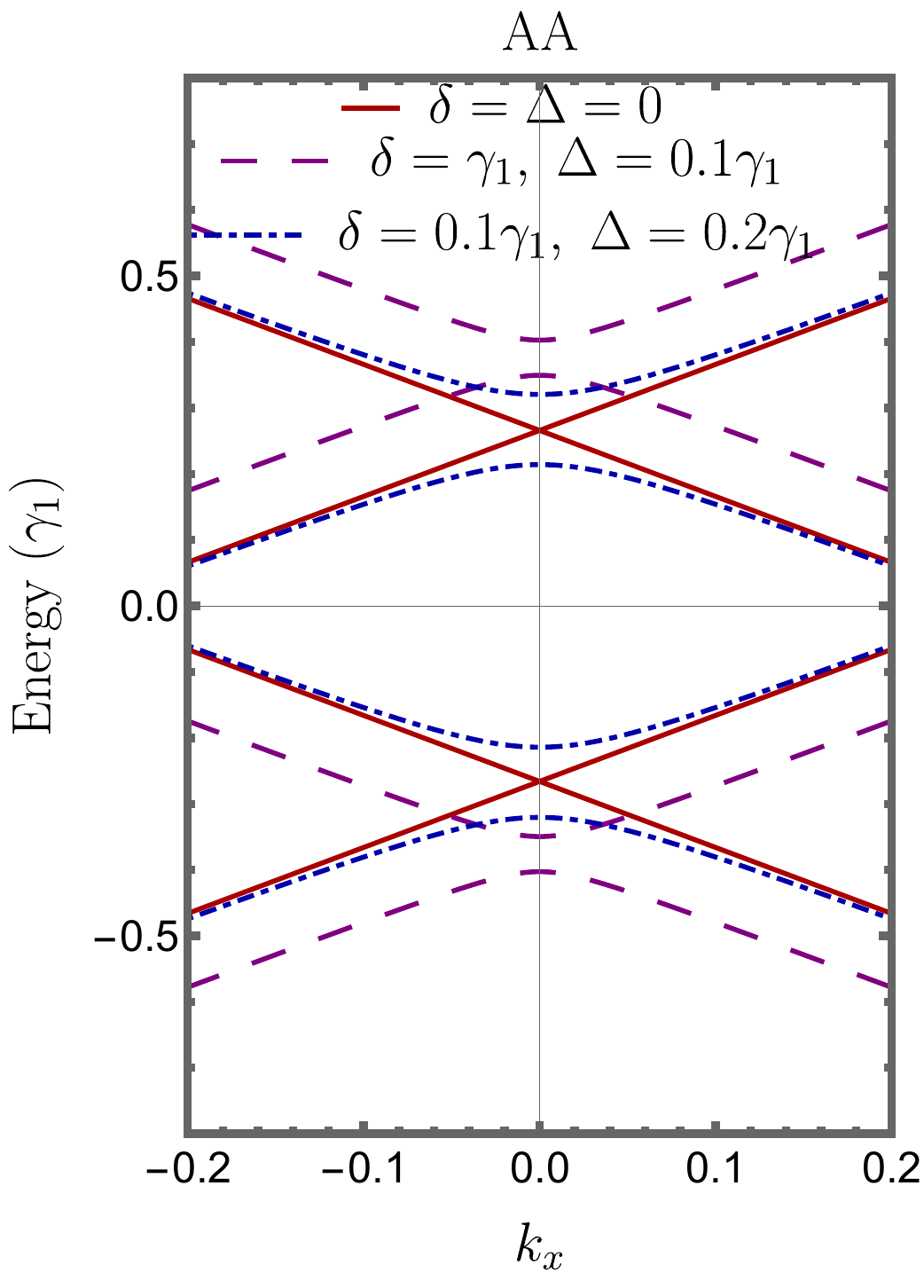}}\subfloat[]{\includegraphics[width=0.42\linewidth]{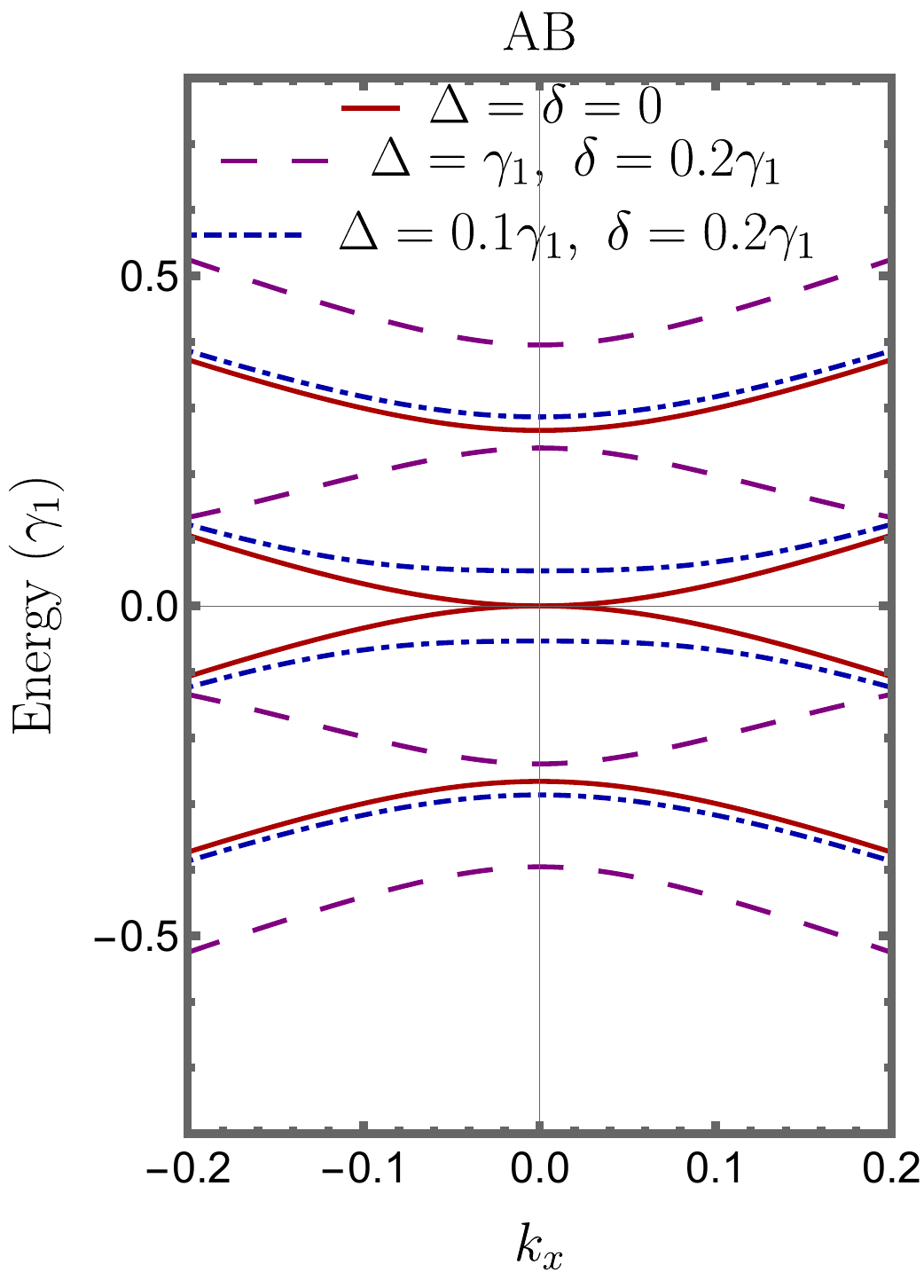}}
\caption{(Color online) Two different geometries for bilayer and single graphene layer interfaces. (a) AA or AB  bilayer graphene  sandwiched between  gapped single layer garphene:  ($\text{SLG}+\Delta/\text{AA-BLG}/\text{SLG}+\Delta$, $\text{SLG}+\Delta/\text{AB-BLG}/\text{SLG}+\Delta$). (b) Gapped AA or gapped AB  bilayer graphene sandwiched between pristine single layer garphene $\text{SLG}/\text{AA-BLG}+(\Delta, \delta)/\text{SLG}$, and $\text{SLG}/\text{AB-BLG}+(\Delta, \delta)/\text{SLG}$. Energy spectrum  for AA (c) and AB (d) for various values of $\delta$ and $\Delta$.
}\label{AA-Energy}
\end{figure}

We study the transport properties of  SLG, AA-BLG, and AB-BLG junctions that can be created from the blocks shown in Fig. \ref{AA-Energy}. Assume an electron with energy $E$ goes from left to right, then the first configuration considered here consists of a gapped SLG while in between they are connected into an AB-BLG or AA-BLG stack. We will call it  $\text{SLG}+\Delta/\text{AA-BLG}/\text{SLG}+\Delta$, and $\text{SLG}+\Delta/\text{AB-BLG}/\text{SLG}+\Delta$, as shown  in Fig. \ref{AA-Energy}~(a). The second configuration consists of a pristine SLG while in between they are connected to a gapped and biased AB-BLG or AA-BLG stack. This configuration will be referred to as $\text{SLG}/\text{AA-BLG}+(\Delta, \delta)/\text{SLG}$, and $\text{SLG}/\text{AB-BLG}+(\Delta, \delta)/\text{SLG}$, as shown in  Fig. \ref{AA-Energy}~(b). In  \cite{gonzalez2010electronic}  studies, they examined similar configurations in pristine cases. Our findings are consistent with their results within the respective limits, but we extend them by taking into account the presence of an energy gap. This will be accomplished by taking into account $\Delta$ in SLG and ($\Delta$, $\delta$) in BLG for a range of Fermi energy and bilayer region  length $d$. Furthermore, we calculate the conductance as a function of the energy gap. This is important since it shows new results and also the experimentally observable signatures of the new physics, like measuring the interlayer hopping in bilayer graphene. Our main results can be summarized as follows:

For $\text{SLG}+\Delta/\text{AA-BLG}/\text{SLG}+\Delta$ the transmission throughout the system shows anti-resonance due to interference from two scattered channels. The conductance as a function of the length  $d$ oscillates with two distinct periods at a fixed energy $E=0.5\gamma_1$ and exhibits anti-Klein tunneling due to the presence of $\Delta$. Now, fixing the energy to $E=2\gamma_1$, the conductance shows one obvious phase. The conductance reveals two distinct patterns of anti-resonances with identical periods when the energy is slightly increased to $2.5\gamma_1$. 
The conductance oscillates for 
 $\text{SLG}+\Delta/\text{AB-BLG}/\text{SLG}+\Delta$ due to finite-size influences. When, $E=0.5\gamma_1$ there is no zero-conductance compared to the $\text{SLG}+\Delta/\text{AA-BLG}/\text{SLG}+\Delta$ case, but for ($E=2\gamma_1$, $E=2.5\gamma_1$) we get anti-resonances with zero conductance. The conductance has different minima as a function of the Fermi energy for each value of the band gap $\Delta$, whereas for $E>\gamma_1$, the minima of the conductance match for all $\Delta$ values. As a result, the conductance is determined by the length $d$  of  the bilayer region.
As a function of the band gap, the results show that $G$ has the same minima regardless of the $d$ value in $\text{SLG}+\Delta/\text{AA-BLG}/\text{SLG}+\Delta$, and $d$ remains relevant in $\text{SLG}+\Delta/\text{AB-BLG}/\text{SLG}+\Delta$ $d$, resulting in a global energy gap.
We show how the band gap and bias affect conductance as a function of $d$ in the second configuration $\text{SLG}/(\text{AA-BLG $\&$ AB-BLG})+(\Delta, \delta)/\text{SLG}$ over a wide range of Fermi energies. 
We observe the appearance of peaks in the conductance profile with different periods and shapes, and also the presence of Klein tunneling with zero-conductance in contrast to the first configuration. For lengths greater than $10$, an oscillating dependence appears. However, the conductance disappears and shows anti-Klein tunneling when $d$ is smaller than $10$.
We investigate the conductance as a function of the bias $\delta$. For $\text{SLG/AA-BLG}+(\Delta,~\delta)/\text{SLG}$, the results show anti-resonances for $\delta<0.1\gamma_1$. Furthermore, we observe the conductance steadily decreasing for a large value of the bias regardless of the length of the bilayer region.  
 For $\Delta=0.5\gamma_1$, we get zero conductance with extra peaks in the conductance profile attributed to the transmitting channels in the AA-BLG. The findings show that the conductance reflects opposite behavior as a function of the interlayer bias and more specifically, the Fermi energy. On the other hand, the conductance of $\text{SLG/AB-BLG}+(\Delta,~\delta)/\text{SLG}$ configuration has distinct characteristics. It begins to conduct with maxima when $ E = 0.5\gamma_1$ and minima starting from $ E > 0.5\gamma_1$. 

The paper under discussion is organized as follows. In Sec. \ref{ELECTRONIC MODEL}, we introduce the full-band continuum model used to describe both configurations. Additionally, we discuss conductance calculations, for which we use the wavefunction correspondence technique. Numerical findings and a discussion of conductance are discussed in Section \ref{Numerical Results AND DISCUSSION}. Finally, in Sec. \ref{Conclusion}, we summarize our main conclusions. 
\section{ELECTRONIC MODEL}
\label{ELECTRONIC MODEL}

\subsection{AA stacking}
Bilayer graphene (BLG) consists of two stacked hexagonal monolayers (MLG). As we refer to them, $A_1$ and $B_1$ for the layer $1$, $A_2$ and $B_2$ for the layer $2$. For AA stacking, both sublattices of layer $1$ are placed directly on top of the two sublattices $A_2$ and $B_2$ of layer $2$. The intralayer interatomic distance is $a_0 = 1.42  \textup{~\AA}$, and it is proportional to the lattice constant  $a =\sqrt{3}a_0$ and the interplanar spacing $c_{\text{AA}} = 3.55 \textup{~\AA}$. In the wave function basis $\psi(x,y)=
\left[\psi_{A_{1}}(x,y),\psi_{B_{1}}(x,y),\psi_{A_{2}}(x,y),\psi_{B_{2}}(x,y)\right]^{\dagger}$, with $\dagger$ represents the row vector's transpose, the Hamiltonian,  near  the valley  $\textbf{K}$, can be expressed as \cite{neto2009electronic, gonzalez2010electronic, brey2006electronic, brey2006edge}
\begin{equation}\label{1}
\mathcal{H}_{AA}=\begin{pmatrix}
0 & v_{F}\pi^{\dag} & \gamma_{1} & 0 \\
v_{F}\pi & 0 &0 &  \gamma_{1}\\
\gamma_{1} & 0 &0 & v_{F}\pi^{\dag} \\
0 &  \gamma_{1} &  v_{F}\pi&0 \\
\end{pmatrix},
\end{equation}
where $v_{F}=\frac{\gamma_{0}}{\hbar}\frac{3a_0}{2}\approx 10^{6}$ m/s is the Fermi velocity for electrons in each graphene layer, $\gamma_0$ is the intralayer coupling between  atoms,  with $\pi=p_{x}+i p_{y}$ are the in-plan momenta and its conjugate with $p_{x,y}=-i\hbar\partial_{x,y}$.$\gamma_1$ is the  interlayer coupling term, in our case we take just $\gamma_1=0.266$ eV, in conformity with the experimental findings \cite{ohta2006controlling, malard2007probing}. Despite the fact that the other interlayer terms have a small impact on transmission \cite{van2013four}. Two different interactions will be introduced to pristine AA-BLG, both referring to an energy gap \cite{mccann2013electronic,predin2016trigonal}  a
band gap $\Delta$, and interlayer bias term $\delta$ given by
\begin{align}
&\begin{gathered}
\mathcal{H}_\Delta=\operatorname{Diag}\{\Delta,-\Delta, \Delta, -\Delta\},
\end{gathered}\\
&
\begin{gathered}
\mathcal{H}_\delta=\operatorname{Diag}\{\delta,\delta, -\delta, -\delta\},
\end{gathered}
\end{align}
in which the total Hamiltonian becomes
\begin{equation}
\mathcal{H}=\mathcal{H}_{AA}+\mathcal{H}_\Delta+\mathcal{H}_\delta.
\end{equation}

Because momentum in the $y$ direction is conserved due to translational invariance in that direction, we decompose the spinor as 
\begin{equation}
\psi(x,y)=e^{ik_{y}y}\left[\phi_{A_1},\phi_{B_1},\phi_{A_2},\phi_{B_2}\right].
\end{equation}
Then, using the Schrodinger equation $\mathcal{H}\psi=E\psi$, we get four linked differential equations
\begin{align}
\label{setofequation1}&-i\left[\partial_{x}+k_{y}\right]\phi_{B_{1}}+\gamma_{1}\phi_{A_{2}} =(E-\Delta-\delta)\phi_{A_{1}},\\
\label{setofequation2}&-i\left[\partial_{x}-k_{y}\right]\phi_{A_{1}}+\gamma_{1}\phi_{B_{2}}=(E+\Delta-\delta)\phi_{B_{1}},\\
\label{setofequation3}&-i\left[\partial_{x}+k_{y}\right]\phi_{B_{2}}+\gamma_{1}\phi_{A_{1}} =(E-\Delta+\delta)\phi_{A_{2}},\\ 
\label{setofequation4}&-i\left[\partial_{x}-k_{y}\right]\phi_{A_{2}}+\gamma_{1}\phi_{B_{1}} =(E+\Delta+\delta)\phi_{B_{2}}.
\end{align}
The set (\ref{setofequation1}-\ref{setofequation4})  
can be simplified to a single second-order differential equation for $\phi_{B_{1}}$, as shown below
\begin{equation}\label{second-order-diff-equation}
\left[\partial_{x}^{2}+(k^{\pm}_{x})^{2}\right]\phi_{B_{1}}=0,
\end{equation}
with the wave vectors
\begin{equation}\label{2-modes}
k^{\pm}=\sqrt{-k_{y}^{2}+ \varepsilon^{2}+\beta^2+\gamma_{1}^2\pm 2\varepsilon\lambda},
\end{equation}
where $\beta=\sqrt{\delta^{2}-\Delta^{2}}$, and  $\lambda=\sqrt{\gamma_{1}^{2}+\delta^{2}}$. The energy spectrum for the system is provided by Eq. (\ref{2-modes}), which follows that
\begin{equation}
E_{s'}^{s}=s'\sqrt{k^{2}+\tau^2+\gamma_{1}^{2}+s\sqrt{\lambda(k^2+\Delta^2)}},
\end{equation}
where $s,s'=\pm$, $\tau=\sqrt{\delta^{2}+\Delta^{2}}$, and $k=\sqrt{k_{x}^{2}+k_{y}^{2}}$.

\subsection{AB stacking}

In the case of AB-BLG, we considered the Bernal stacking \cite{bernal1924structure}, which means that the two graphene layers are constructed in such a way that the sublattice $A_1$ is precisely on top of the sublattice $B_2$. The associated Hamiltonian  is given by
\begin{equation}
\mathcal{H}=\mathcal{H}_{AB}+\mathcal{H}_\Delta+\mathcal{H}_\delta.
\end{equation}
Following the same steps as AA-stacking,  we find the energy spectrum  of AB-BLG  as  
\small\begin{equation}
E_{s'}^{s}=s'\left[k^{2}+\tau^2+\frac{\gamma_{1}^{2}}{2}+s\sqrt{k^{2}\left(\gamma_{1}^{2}+4\delta^{2}\right)+\left(\frac{\gamma_{1}^{2}}{2}+2\delta\Delta\right)^{2}}\right]^{\frac{1}{2}}.
\end{equation}
For the corresponding propagating modes in AB-BLG, there are two labeled by $k^{+}$ and $k^{-}$. They are
\begin{equation}
k^{\pm}=\sqrt{-k_{y}^{2}+ E^{2}+\beta\pm\sqrt{E^{2}(\gamma_{1}^{2}+4\delta^{2})-\gamma_{1}^{2}(\delta-\Delta)^2}}.
\end{equation}

In Fig. \ref{AA-Energy} (c,d) we illustrate the energy spectra of the AA-BLG and AB-BLG for various system model parameters. We see that for biased $\delta$ and gapped $\Delta$ AA-BLG, the two Dirac cones are  shifted at the two pairs of values ($\Delta=\gamma_1$, $\delta=0.1\gamma_1$) (purple dashed line)  and  ($\Delta=0.1\gamma_1$, $\delta=0.3\gamma_1$) (blue dot dashed line), whereas   the proportionality to $\gamma_1$ create parabolic cones.
The two bands are flipped and located at $\pm\sqrt{\gamma^2_{1}+\beta^2-2\delta\Delta}$ for the biased  and gapped AB-BLG, and the touching bands are moved apart by $\delta+\Delta$. 

The transmission, reflection, and the coefficients of the
wave functions in the  AA-BLG or AB-BLG  part are identified by implementing the required boundary conditions starting with $x=0$ and finally with $x=d$ for both regions of  AA-BLG or AB-BLG, see Appendix \ref{Appendix-D}. Using the transmission probabilities, we can calculate the conductance as a function of the energy given by the Landauer-Buttiker formula
\begin{equation}
G(E)=G_{0}T(E),
\end{equation}
where $G_{0}=2e^{2}/h$.

\section{Numerical Results AND DISCUSSION}
\label{Numerical Results AND DISCUSSION}

\subsubsection{$\text{SLG}+\Delta/\text{AA-BLG $\& $ AB-BLG}/\text{SLG}+\Delta$}
\begin{figure*}[htp]
	\centering
	\includegraphics[width=0.9\linewidth]{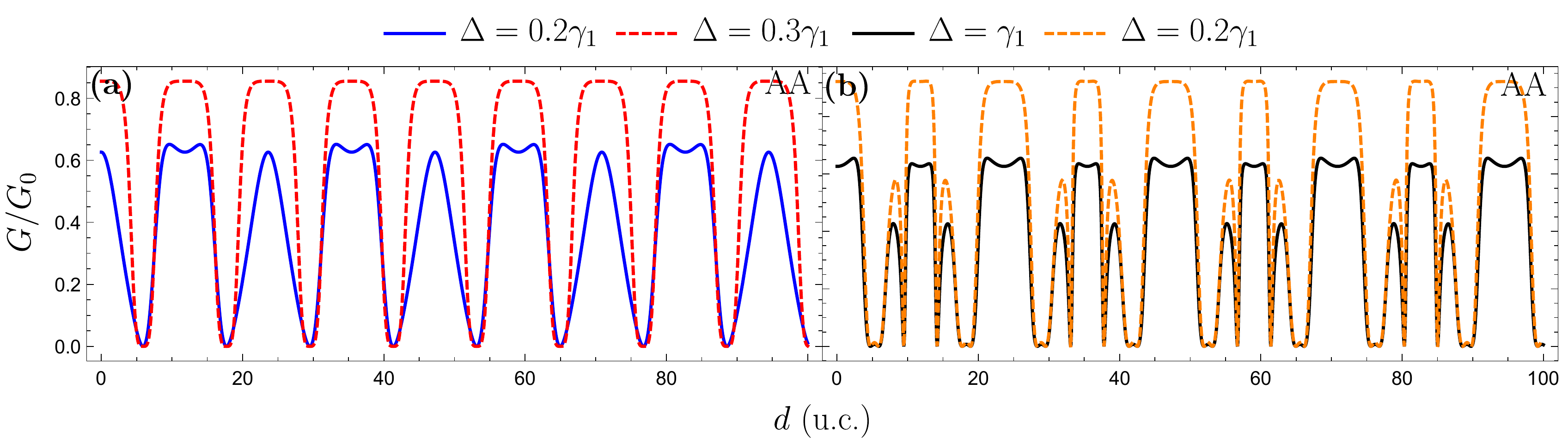}\\\includegraphics[width=0.9\linewidth]{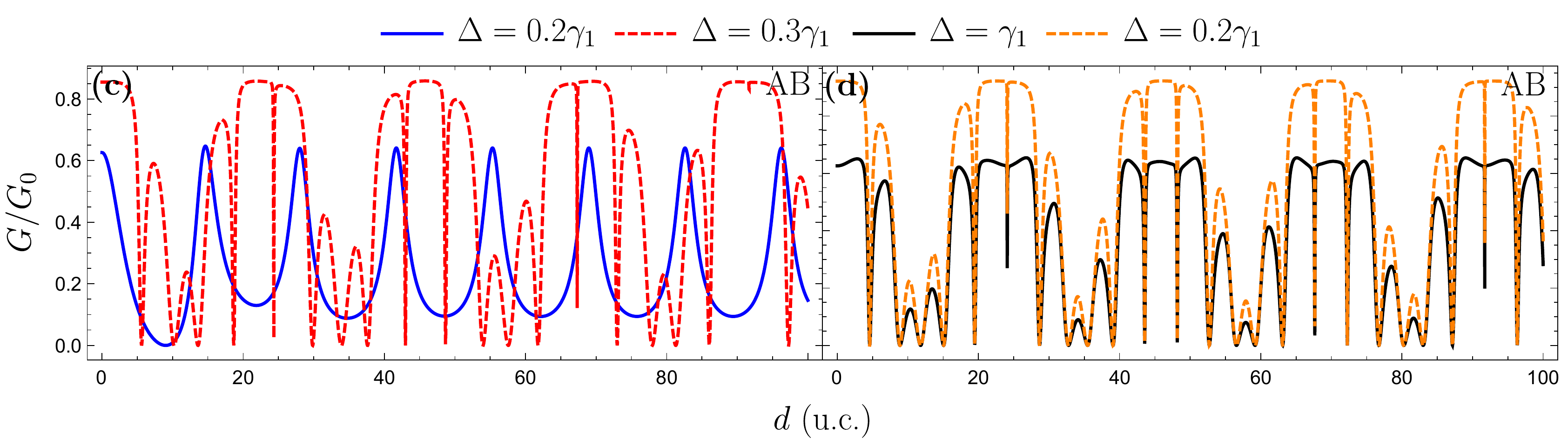}	
	\caption{(Color online) {Conductance as a function of bilayer region length $d$ through $\text{SLG}+\Delta/\text{AA-BLG}/\text{SLG}+\Delta$ (top panel) and $\text{SLG}+\Delta/\text{AB-BLG}/\text{SLG}+\Delta$ (bottom panel)  junctions, with energies $E=0.5\gamma_1$ (blue solid lines), $E=2\gamma_1$ (red dashed lines), and $E=2.5\gamma_1$ (black solid lines and orange solid lines)}.}\label{FigAA-AB-Delta}
\end{figure*}
\begin{figure*}
	\centering
	\includegraphics[width=0.5\linewidth]{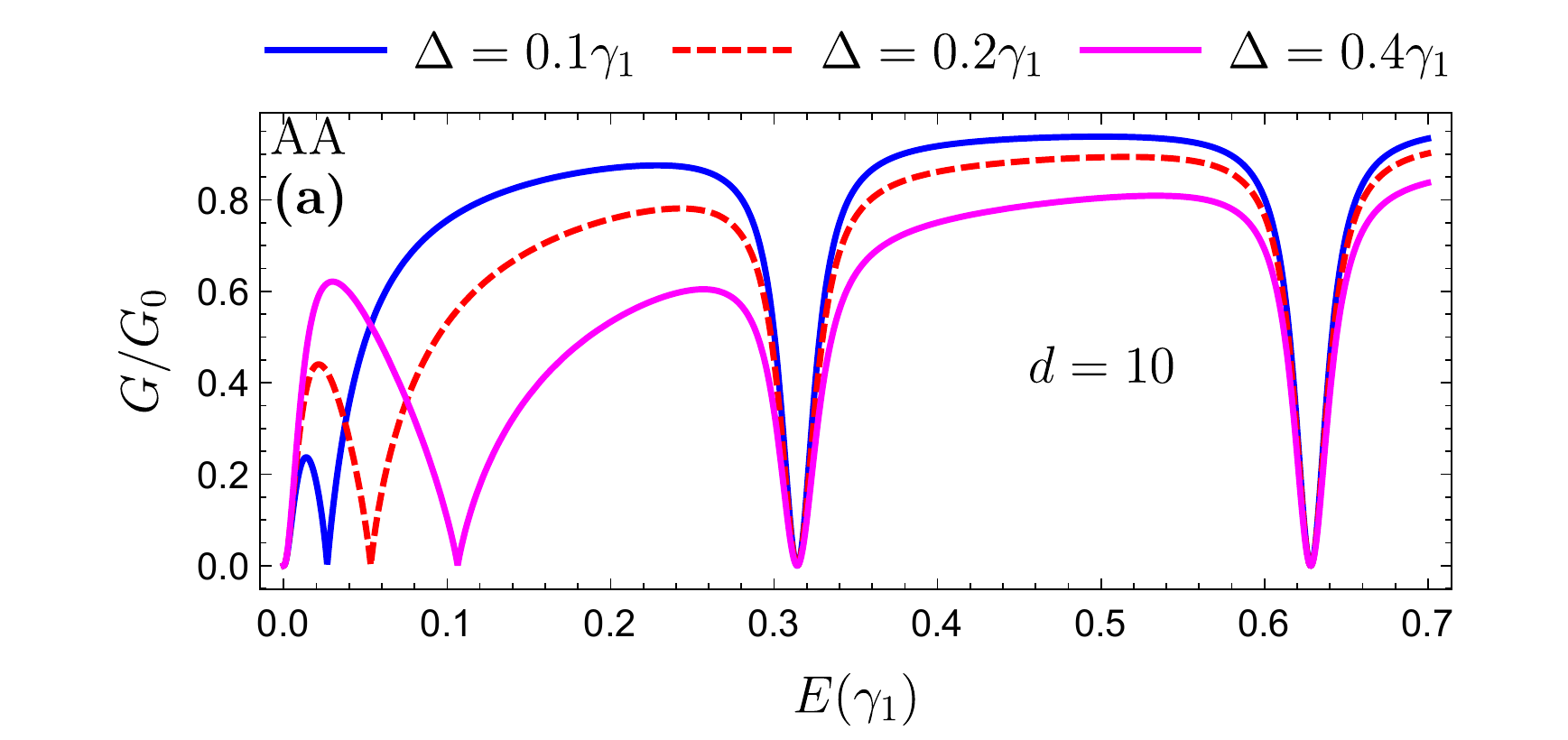}\includegraphics[width=0.5\linewidth]{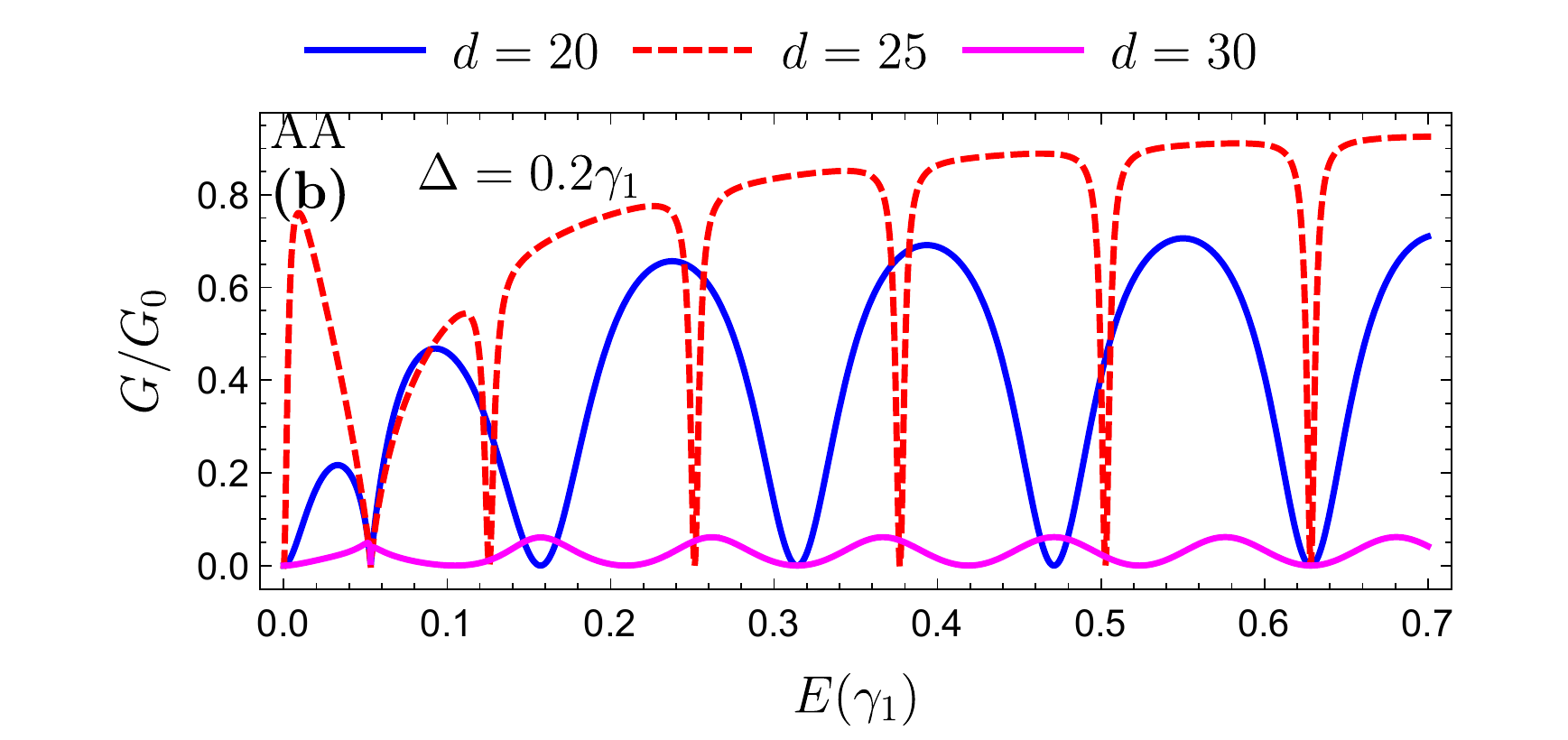}\\\includegraphics[width=0.5\linewidth]{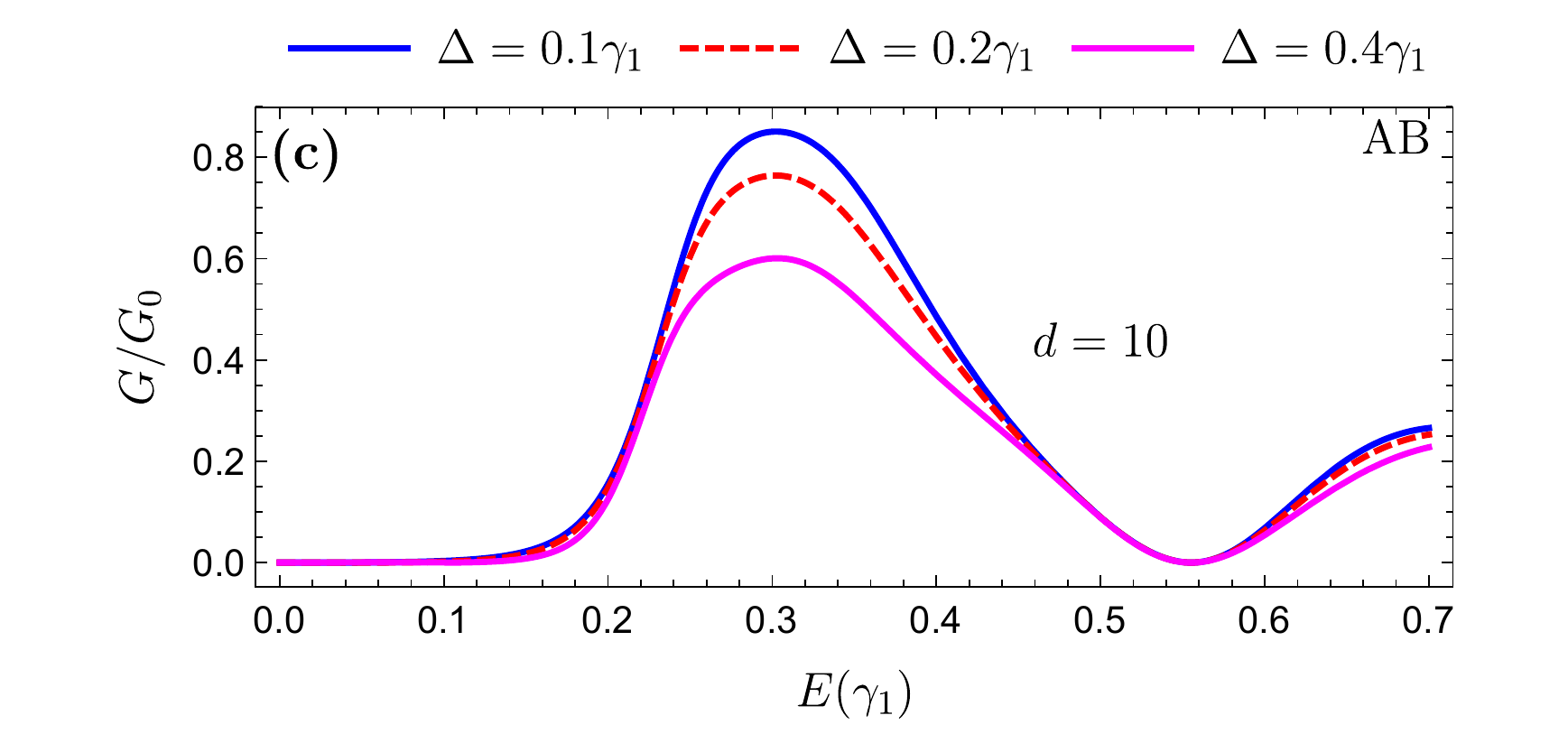}\includegraphics[width=0.5\linewidth]{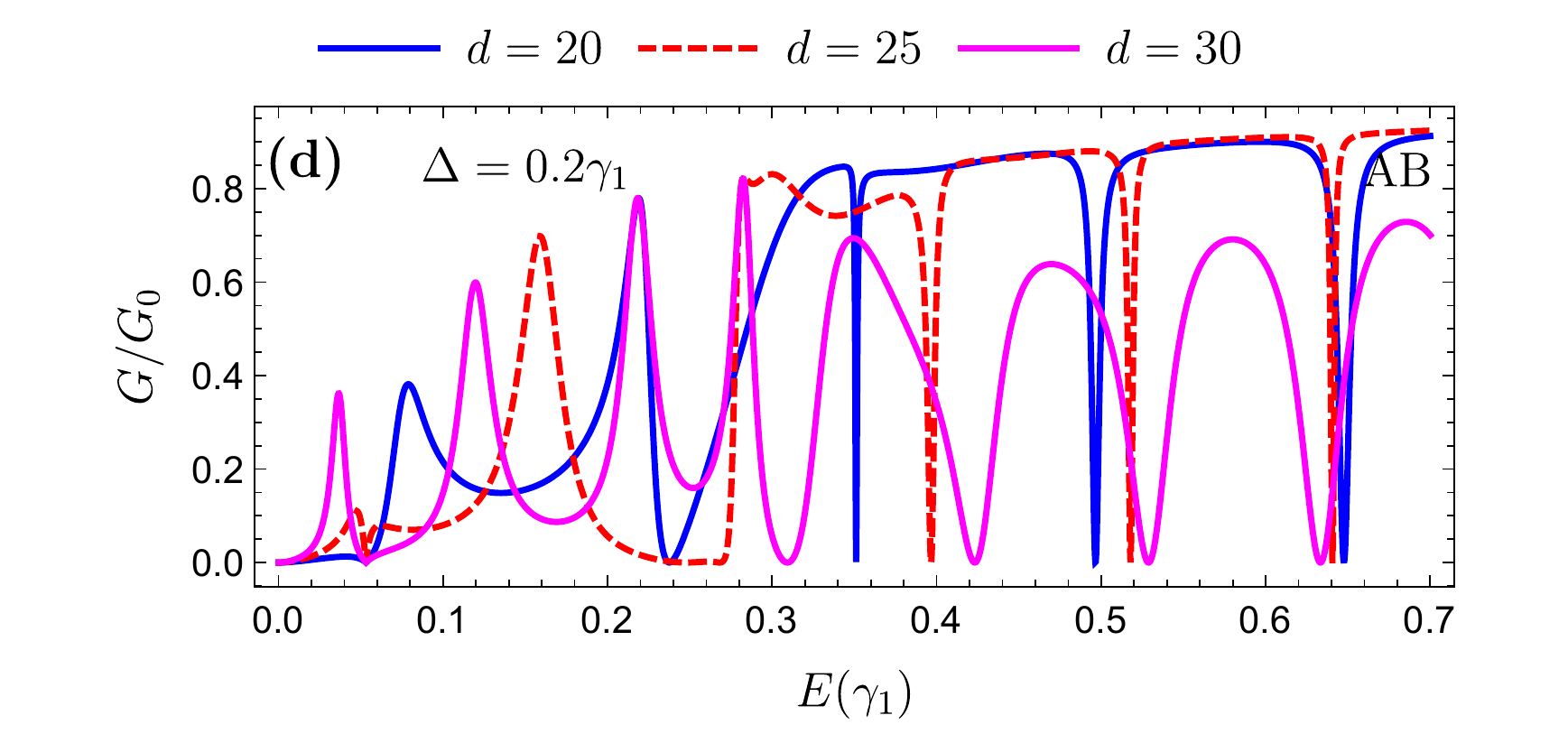}	
	\caption{(Color online) {Conductance as a function of the Fermi energy through $\text{SLG}+\Delta/\text{AA-BLG}/\text{SLG}+\Delta$, and $\text{SLG}+\Delta/\text{AB-BLG}/\text{SLG}+\Delta$  junctions.}}\label{fig-conductance-energy1}
\end{figure*}
\begin{figure}[htp]
	\centering
	\includegraphics[width=0.5\linewidth]{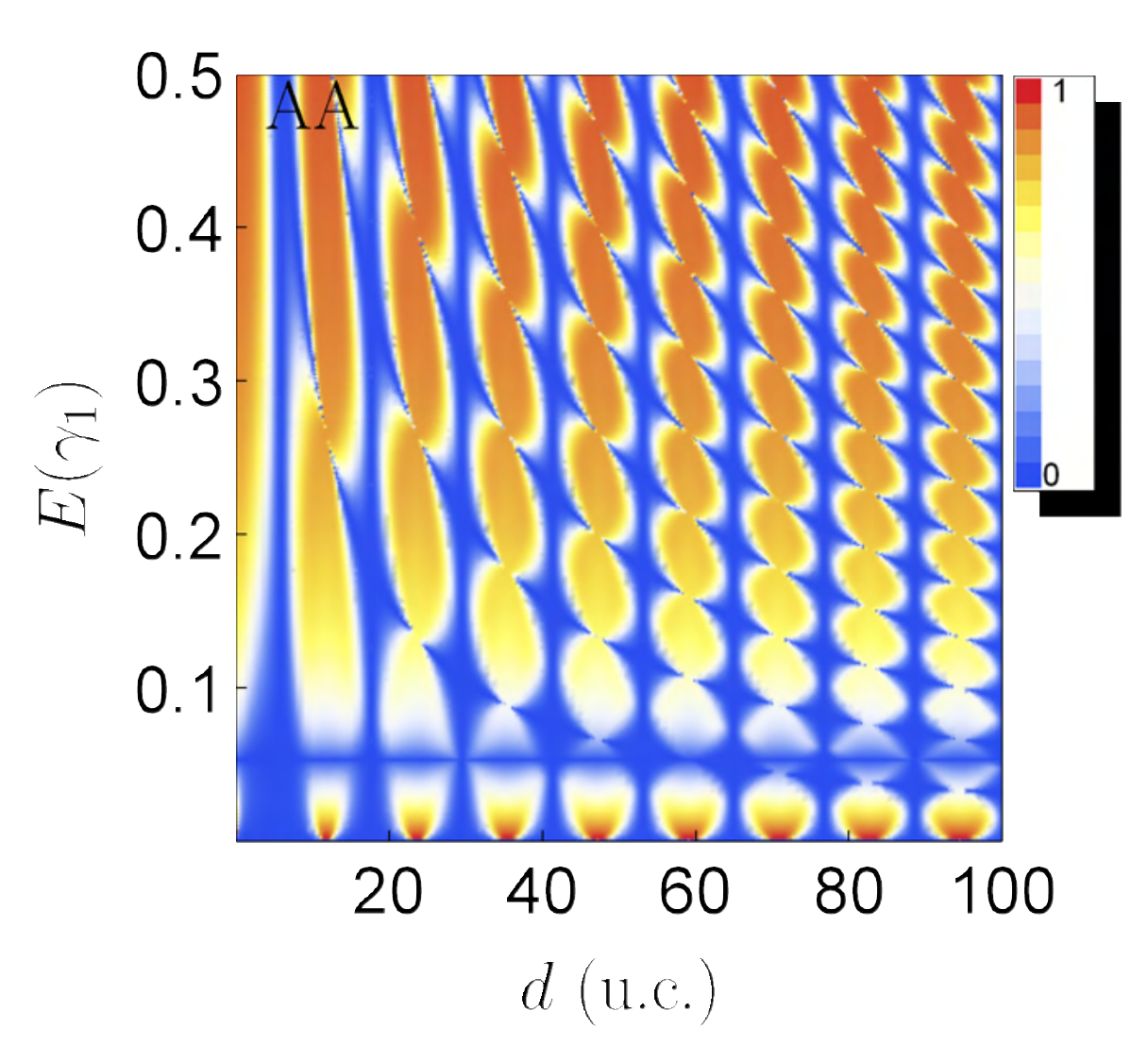}\includegraphics[width=0.5\linewidth]{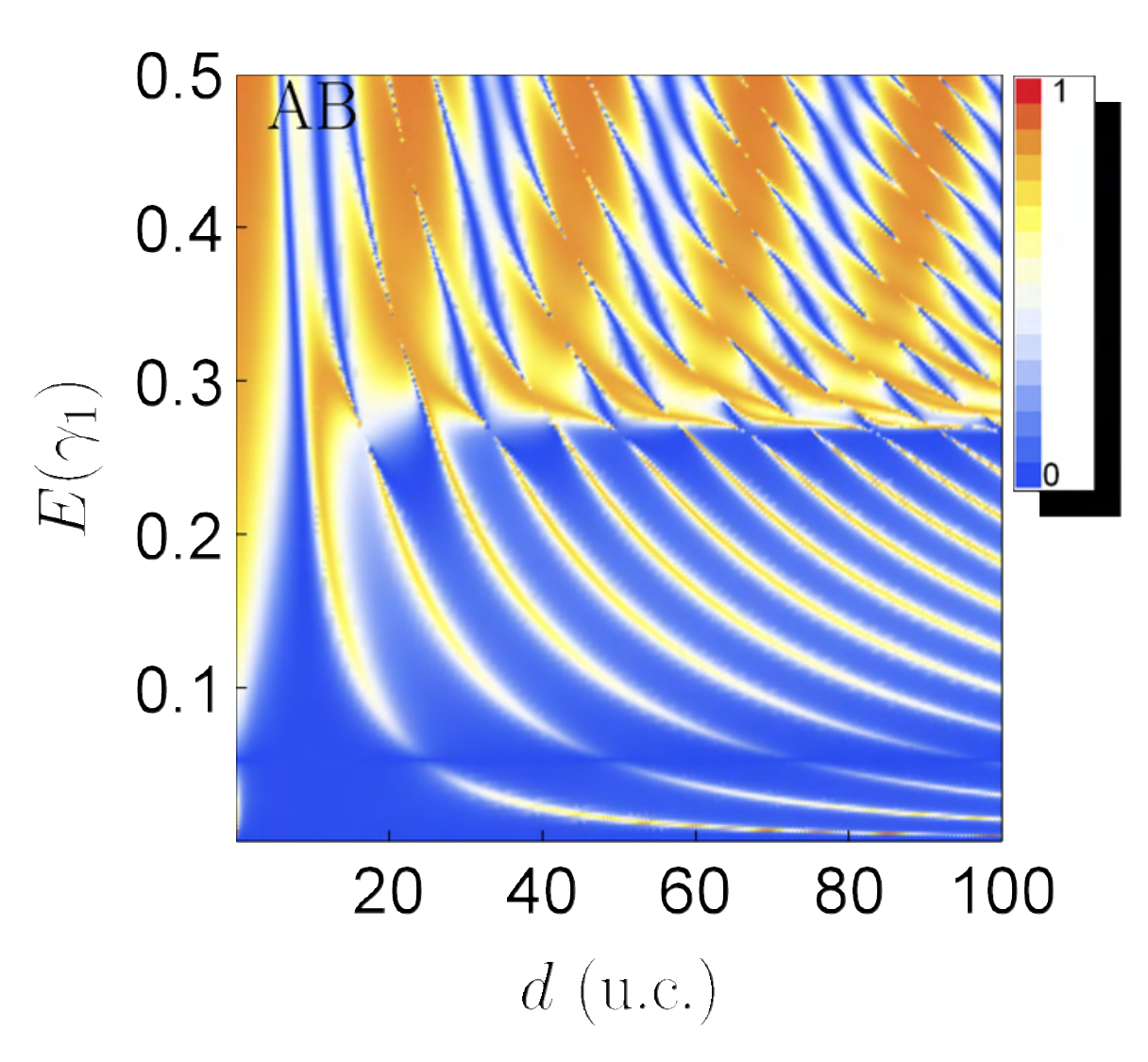}
	\caption{(Color online) {Density plot of the transmission probability as a function of bilayer region  length $d$, and  Fermi energy $E$ through $\text{SLG}+\Delta/\text{AA-BLG}/\text{SLG}+\Delta$, and $\text{SLG}+\Delta/\text{AB-BLG}/\text{SLG}+\Delta$, with  $\Delta=0.2\gamma_1$.}}\label{DP1}
\end{figure}
\begin{figure}[htp]
	\centering
	\includegraphics[width=0.5\linewidth]{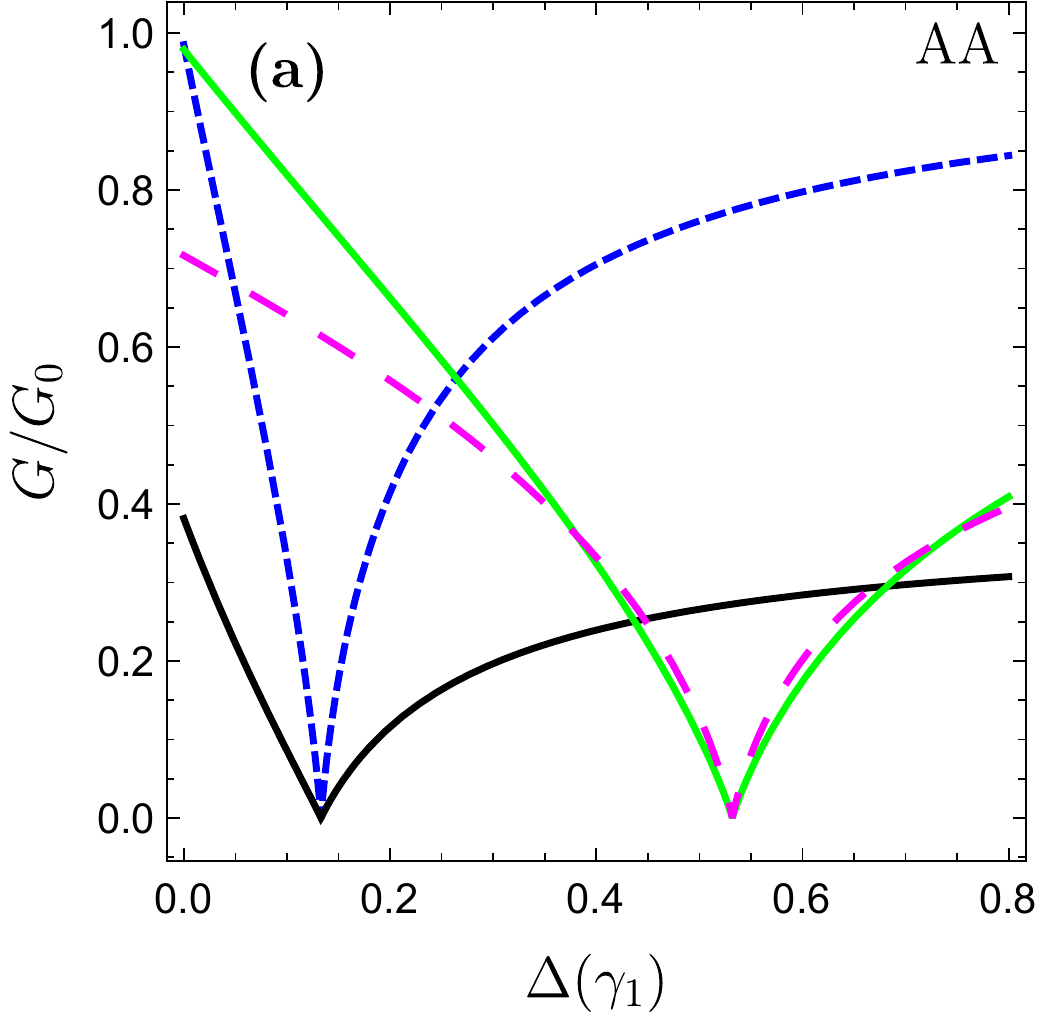}\includegraphics[width=0.5\linewidth]{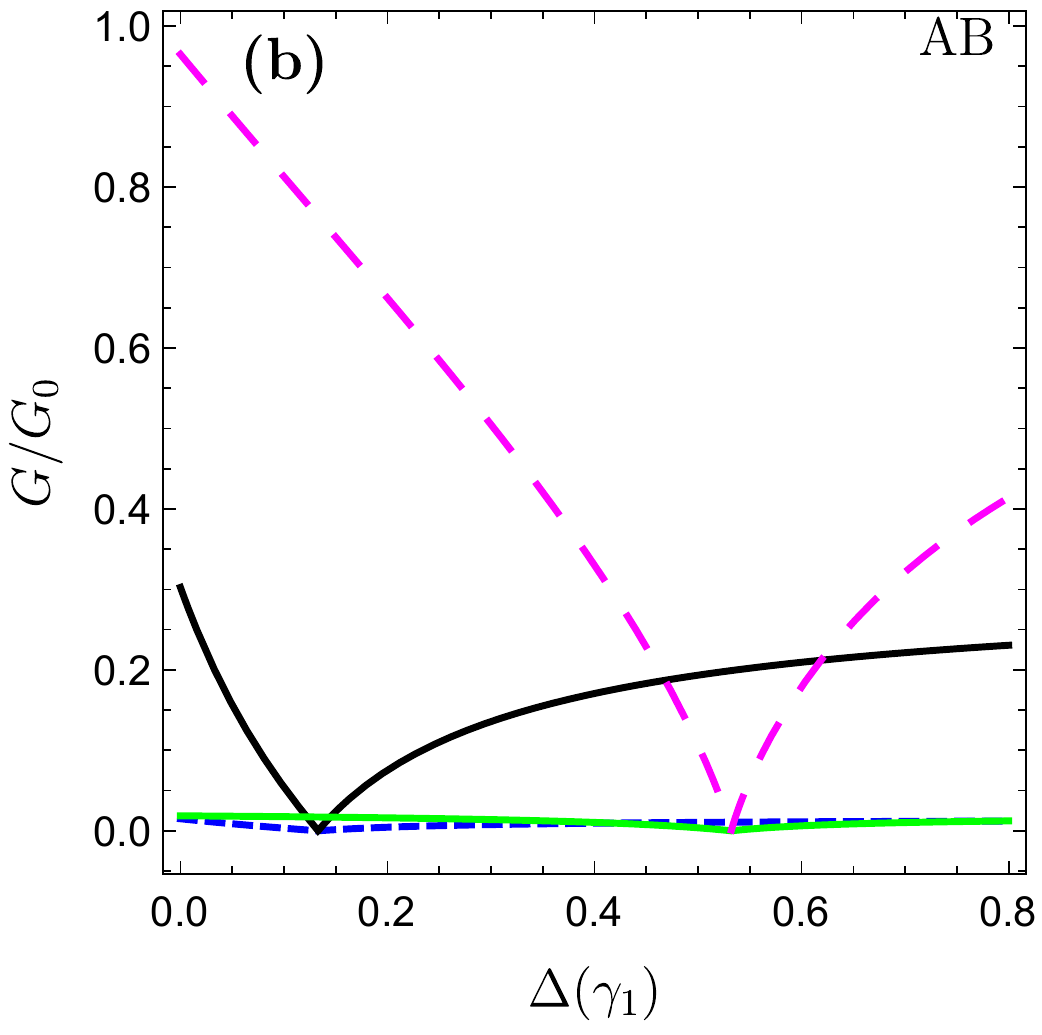}	
	\caption{(Color online) Conductance as a function of the band gap $\Delta$, for $\text{SLG}+\Delta/\text{AA-BLG}/\text{SLG}+\Delta$, and $\text{SLG}+\Delta/\text{AB-BLG}/\text{SLG}+\Delta$ junctions. $(E,d)=(0.5\gamma_1, 10)$ (blue dashed line), $(E,d)=(0.5\gamma_1, 20)$ (black solid line), $(E,d)=(2\gamma_1, 10)$ (magenta dashed line), $(E,d)=(2\gamma_1, 20)$ (blue dashed line). }\label{conductance as function of Delta1}
\end{figure}

We numerically examine and discuss our main findings for gapped SLG with pristine AA-BLG stacking and pristine AB-BLG stacking, namely $\text{SLG}+\Delta/\text{AA-BLG}/\text{SLG}+\Delta$, and $\text{SLG}+\Delta/\text{AB-BLG}/\text{SLG}+\Delta$ configurations. In Fig. \ref{FigAA-AB-Delta}, the conductance is plotted as a function of the length $d$. Here,   $d$ is expressed in the armchair ribbon (u.c.) length, which is $3a_0$ \cite{gonzalez2010electronic}. 
For $\text{SLG}+\Delta/\text{AA-BLG}/\text{SLG}+\Delta$ in Fig.  \ref{FigAA-AB-Delta} (a), the transmission through the system shows anti-resonances due to the interference of two scattered channels. 
At a fixed energy of $E=0.5\gamma_1$ and a band gap of $\Delta=0.2\gamma_1$, the conductance  oscillates with two distinct periods of 24 and 33 u.c. (blue solid line) as a function of length $d$. It also exhibits anti-Klein tunneling due to the presence of $ \Delta $, with one of the periods changing compared to the result obtained in \cite{gonzalez2010electronic}. 
For $E=2\gamma_1$ and $\Delta=0.3\gamma_1$, the conductance shows one obvious phase of 4 u.c. 
As  in Fig. \ref{FigAA-AB-Delta} (b) for $2.5\gamma_1$ with $\Delta=\gamma_1$ (black solid line) and $\Delta=0.2\gamma_1$ (orange dashed line), the conductance reveals two different patterns of anti-resonances with identical periods of 23 u.c. 
We plotted the conductance for $\text{SLG}+\Delta/\text{AB-BLG}/\text{SLG}+\Delta$ as a function of length $d$ in Fig. \ref{FigAA-AB-Delta} (c). For $E=0.5\gamma_1$ and $\Delta=0.2\gamma_1$ (blue solid line), there is just one transmission channel in the AB-BLG. Indeed, there are no anti-resonances in this energy; the conductance oscillates as a consequence of finite-size influences. These are also known as Febry-Pérot resonances \cite{snyman2007ballistic}. In this case, the presence of our band gap $\Delta$ reduces conductance, but there is no zero-conductance as in the  $\text{SLG}+\Delta/\text{AA-BLG}/\text{SLG}+\Delta$ case. 
%
For energies greater than $\gamma_1$ ($E =2\gamma_1$, $E =2.5\gamma_1$), Fig. \ref{FigAA-AB-Delta}~(d), both configurations have two propagation channels. The case here has been updated, and it now shows anti-resonances with zero conductance. In this case, the anti-resonances have periods that depend on $E$, $\gamma_1$, $d$, and the band gap $\Delta$. 
This remarkable distinction against the AA case is due to the fact that all atoms are connected in the AA pattern, but only half of the atoms in the AB case are linked by interlayer hopping as noted in \cite{gonzalez2010electronic}.

The conductance $G(E)$ for the gapped SLG as a function of the Fermi energy is shown in Fig. \ref{fig-conductance-energy1} for the two investigated geometries with $d=10$ on the left panel and $d= (20,25,30)$ on the right panel. Regarding the first $\text{SLG}+\Delta/\text{AA-BLG}/\text{SLG}+\Delta$ case plotted  in Fig. \ref{fig-conductance-energy1} (a), the conductance shows different minima for each value of the band gap for $E<\gamma_1$. In contrast to $E > \gamma_1$, the conductance minima match for all values of $\Delta=(0.1\gamma_1, 0.2\gamma_1, 0.4\gamma_1)$. The $\text{SLG}+\Delta/\text{AB-BLG}/\text{SLG}+\Delta$ case is shown in Fig. \ref{fig-conductance-energy1} (c), and the conductance  is zero in the energy range $(0, \gamma_1)$ for $d=10 $. Fixing now $\delta$ for $0.2\gamma_1$ and prolonging the bilayer region length $ d $, the additional peaks that can be seen in the conductance pattern in $\text{SLG}+\Delta/\text{AA-BLG}/\text{SLG}+\Delta$ are associated with Fano antiresonances with zero conductance, and in the $\text{SLG}+\Delta/\text{AB-BLG}/\text{SLG}+\Delta$ are associated with Fabry-Pérot-resonances.  It is worth noting that the conductance depends on  $d$ and obviously varies from $0$ when $d > 25$ through $\text{SLG}+\Delta/\text{AA-BLG}/\text{SLG}+\Delta$ as shown in Fig. \ref{fig-conductance-energy1} (b).


The density plot of the transmission probability as a function of Fermi energy and a bilayer region of length $d$ for $\Delta=0.2\gamma_1$ at normal incidence ($k_y = 0$) is shown in Fig. \ref{DP1}. There are two distinct energy regimes based on interlayer coupling. There are no anti-resonances in $\text{SLG}+\Delta/\text{AA-BLG}/\text{SLG}+\Delta$ due to the presence of two transmitting modes, as opposed to $\text{SLG}+\Delta/\text{AB-BLG}/\text{SLG}+\Delta$, which has only one propagating mode. The implication is that for various values of $d$, even after the band gap is taken into account, the conductance is zero, regardless of the energy or $ \delta $. We guarantee the fact that this property can be utilized to quantify the interlayer hopping parameter since the period relies primarily on the $\gamma_1$ parameter.


To investigate the effect of the band gap $\Delta$, we show in Fig. \ref{conductance as function of Delta1} the conductance through $\text{SLG}+\Delta/\text{AA-BLG}/\text{SLG}+\Delta$, and  $\text{SLG}+\Delta/\text{AB-BLG}/\text{SLG}+\Delta$ junctions at normal incidence as a function of $\Delta$ for various Fermi energies and bilayer length $d$. For $\text{SLG}+\Delta/\text{AA-BLG}/\text{SLG}+\Delta$ plotted in Fig. \ref{conductance as function of Delta1} (a),  the results show that whatever the value of $d$, the conductance  has the same minima, $0.13\gamma_{1}$ for $E=0.5\gamma_1$ (blue dashed line, black solid line) and $0.53\gamma_1$ for $E=2\gamma_1$ (purple dashed line,  green solid line). The  $\text{SLG}+\Delta/\text{AB-BLG}/\text{SLG}+\Delta$ case is shown in Fig. \ref{conductance as function of Delta1} (b).
When $d\leqslant10 $ and for $E<\gamma_1$ or $E>\gamma_1$,   conductance remains $0$ for all $\Delta$. We conclude that the bilayer region length $d$ remains relevant such that the system creates a global energy gap.


\subsubsection{$\text{SLG}/(\text{AA-BLG $\&$ AB-BLG})+(\Delta, \delta)/\text{SLG}$}

We present charge carriers tunneling through $\text{SLG}/\text{AA-BLG}+(\Delta, \delta)/\text{SLG}$, and $\text{SLG}/\text{AB-BLG}+(\Delta, \delta)/\text{SLG}$ systems.
\begin{figure*}[ht]
	\centering
	\includegraphics[width=0.9\linewidth]{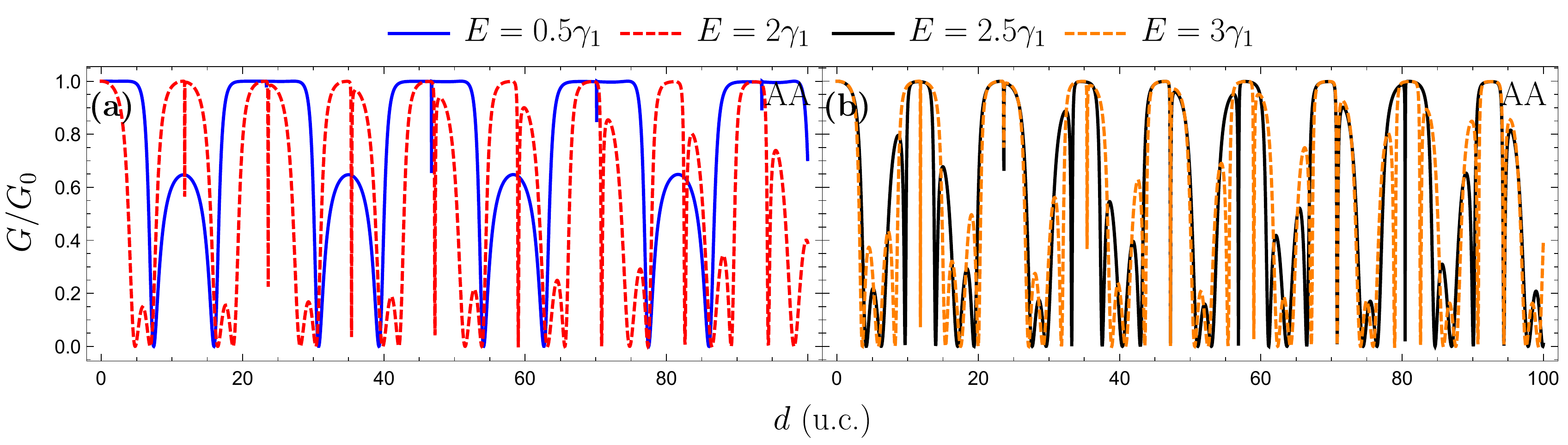}\\\includegraphics[width=0.9\linewidth]{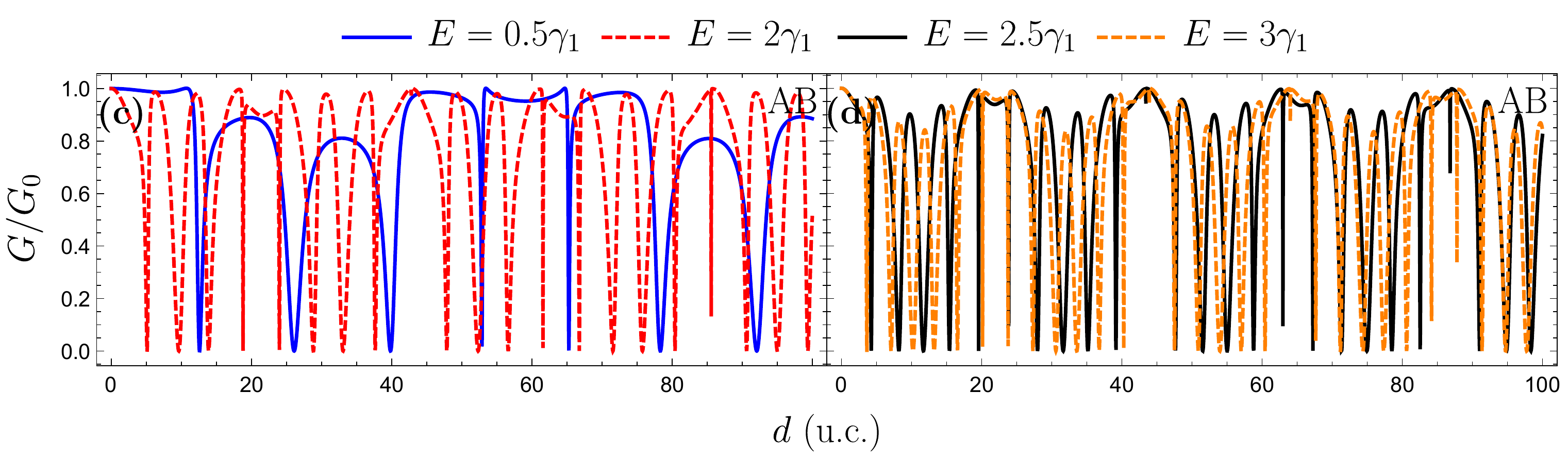}	
	\caption{(Color online) Conductance as a function of bilayer region length $d$  through $\text{SLG}/\text{AA-BLG}+(\Delta, \delta)/\text{SLG}$, and $\text{SLG}/\text{AB-BLG}+(\Delta, \delta)/\text{SLG}$ junctions, with $\delta=0.2\gamma_1$ and $\Delta=0.1\gamma_1$.}\label{FigAA-AB-Delta-delta}
\end{figure*}
The influence of the band gap $\Delta=0.1\gamma_1$ and the bias $\delta=0.2\gamma_1$ on the conductance as a function of the bilayer region of length $d$ for a range of Fermi energy $E= (0.5\gamma_1, 2\gamma_1, 3\gamma_1$) is shown in Fig. \ref{FigAA-AB-Delta-delta}. We see peaks in the conductance profile with different periods and shapes for both geometries $\text{SLG}/\text{AA-BLG}+(\Delta, \delta)/\text{SLG}$ and $\text{SLG}/\text{AA-BLG}+(\Delta, \delta)/\text{SLG}$. The presence of Klein tunneling with zero-conductance in contrast to the first case (see Fig. \ref{FigAA-AB-Delta}). The Klein tunneling becomes attainable if certain parameters are fulfilled, as outlined recently in \cite{van2013klein}. The presence of resonances in the $\text{SLG}/\text{AB-BLG}+(\Delta, \delta)/\text{SLG}$ case can be explained by the presence of charge carriers with different chiralities and the finite size of the AB-BLG, which also corresponds to our recent work \cite{benlakhouy2021transport}. 


\begin{figure*}
	\centering
	\includegraphics[width=0.5\linewidth]{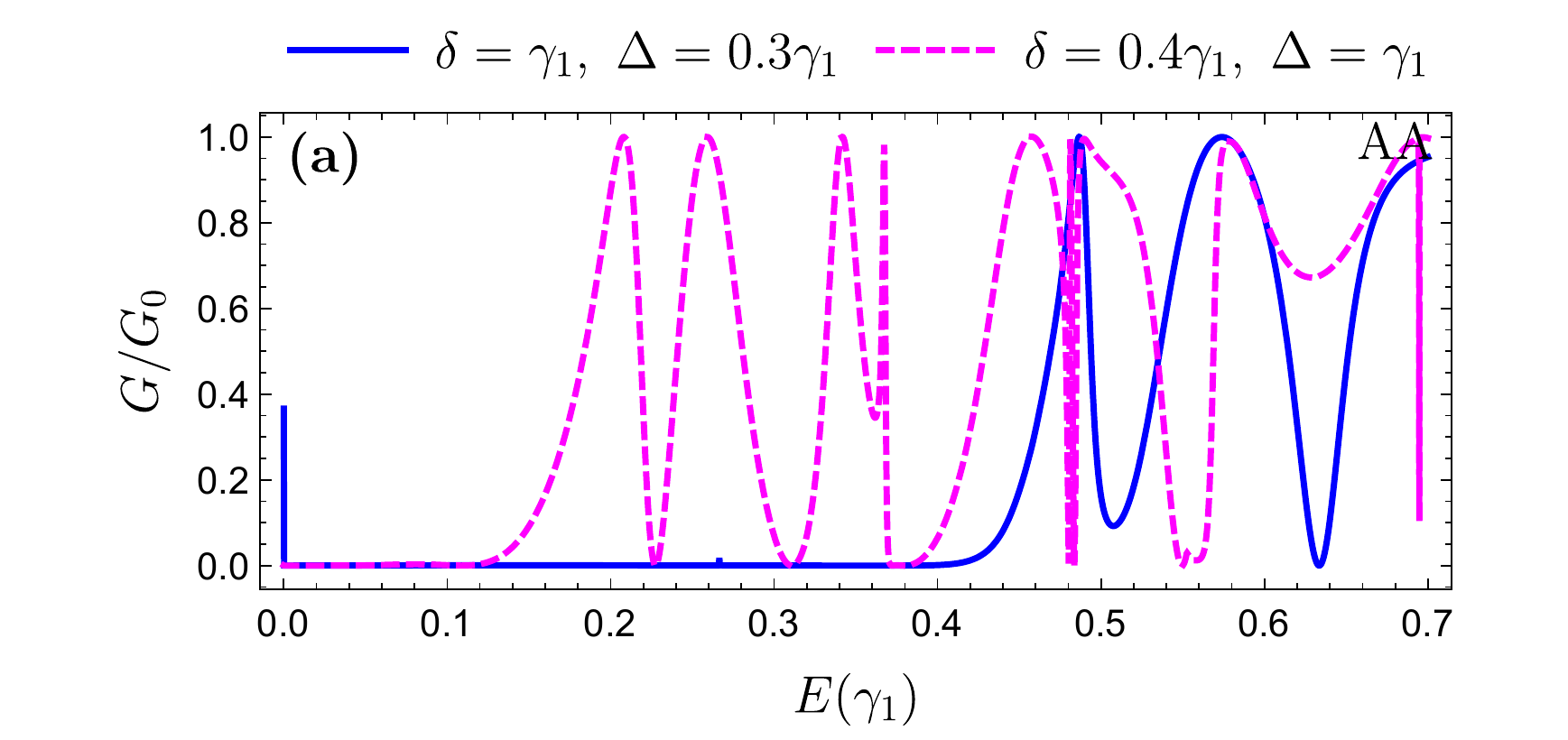}\includegraphics[width=0.5\linewidth]{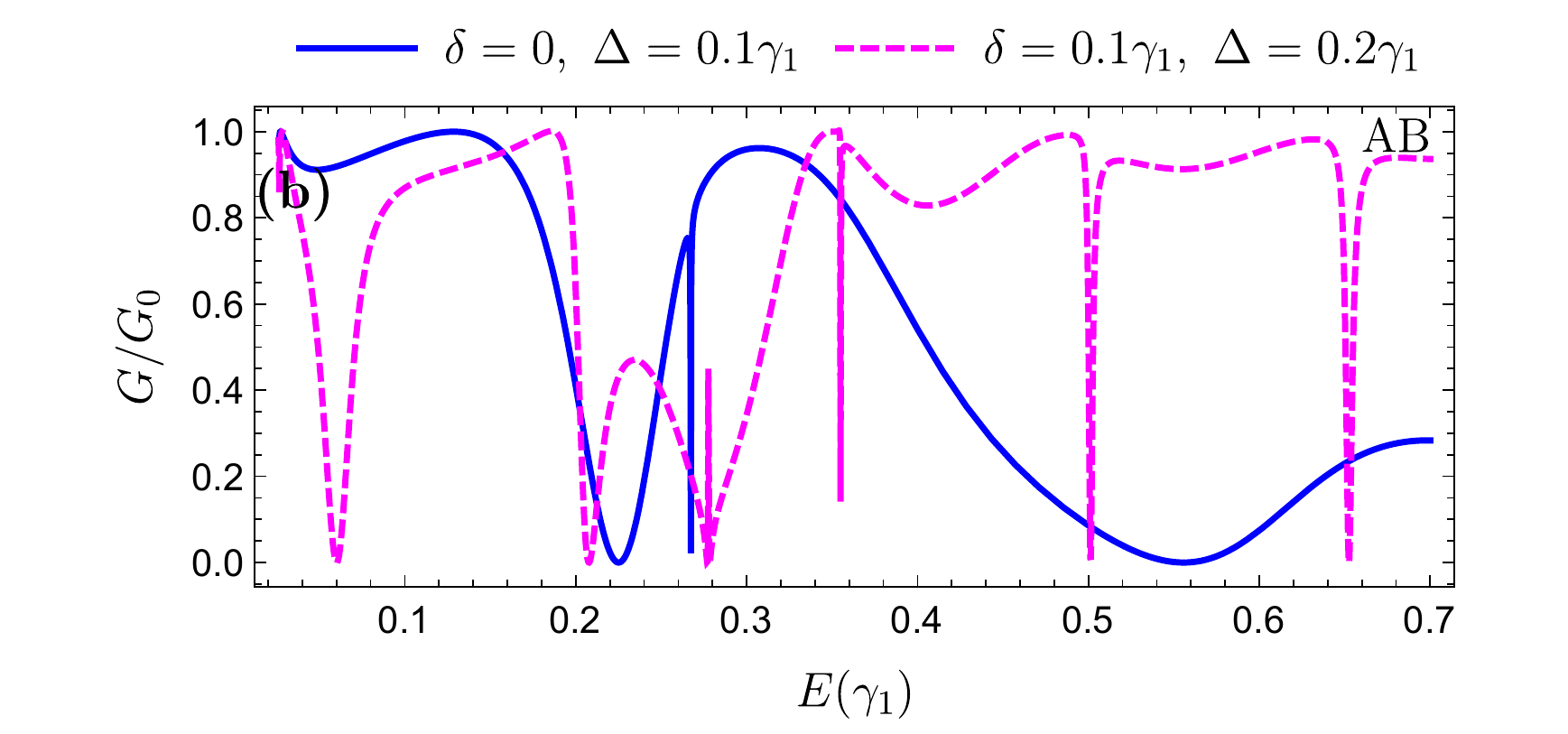}	
	\caption{(Color online) Conductance as a function of the Fermi energy  {through} (a) $\text{SLG}/\text{AA-BLG}+(\Delta, \delta)/\text{SLG}$,  and (b) $\text{SLG}/\text{AB-BLG}+(\Delta, \delta)/\text{SLG}$ junctions, with $d=10$ (blue solid line), and $d=20$ (magenta dashed line).}\label{fig-conductance-energy2}
\end{figure*}
\begin{figure}
	\centering
	\includegraphics[width=0.5\linewidth]{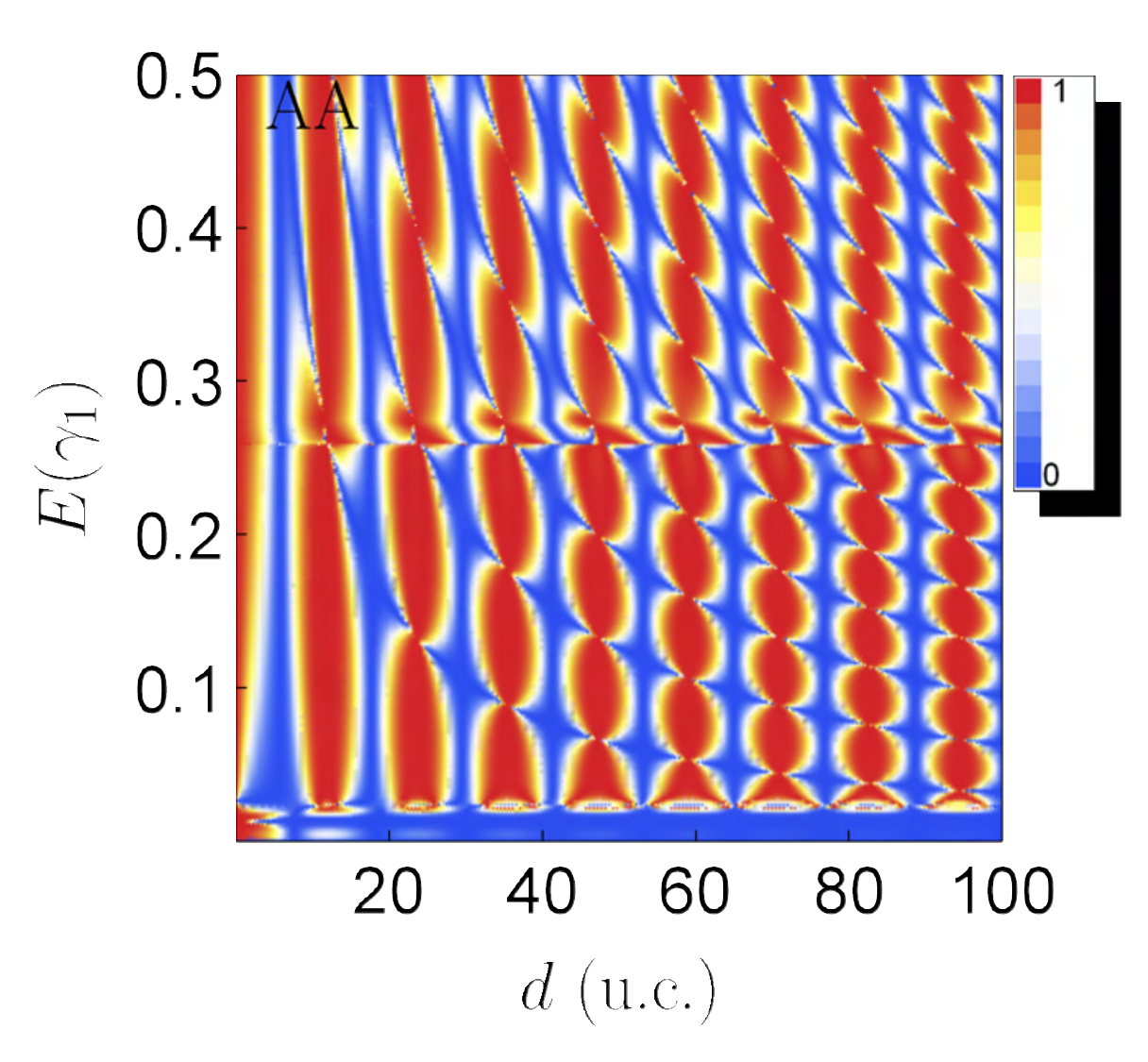}\includegraphics[width=0.5\linewidth]{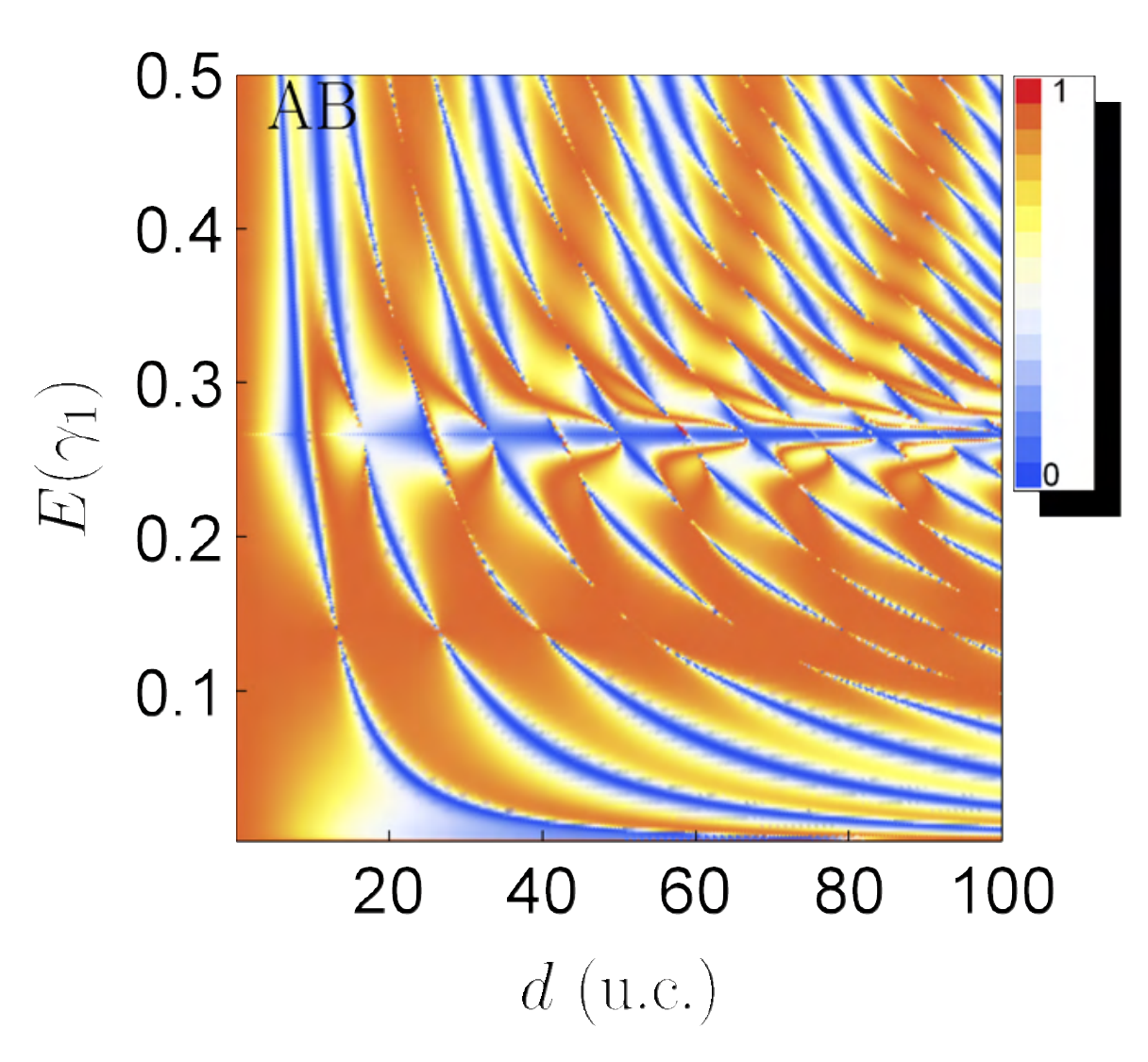}
	\caption{(Color online) Density plot of the transmission  probability as a function of bilayer region of length $d$, and  Fermi energy $E$ {through} $\text{SLG}/\text{AA-BLG}+(\Delta, \delta)/\text{SLG}$, and $\text{SLG}/\text{AB-BLG}+(\Delta, \delta)/\text{SLG}$ junctions, with parameters   $\delta=0.02\gamma_1$, and  $\Delta=0.03\gamma_1$.}\label{DP2}
\end{figure}
\begin{figure*}
	\centering
	\includegraphics[width=0.448\linewidth]{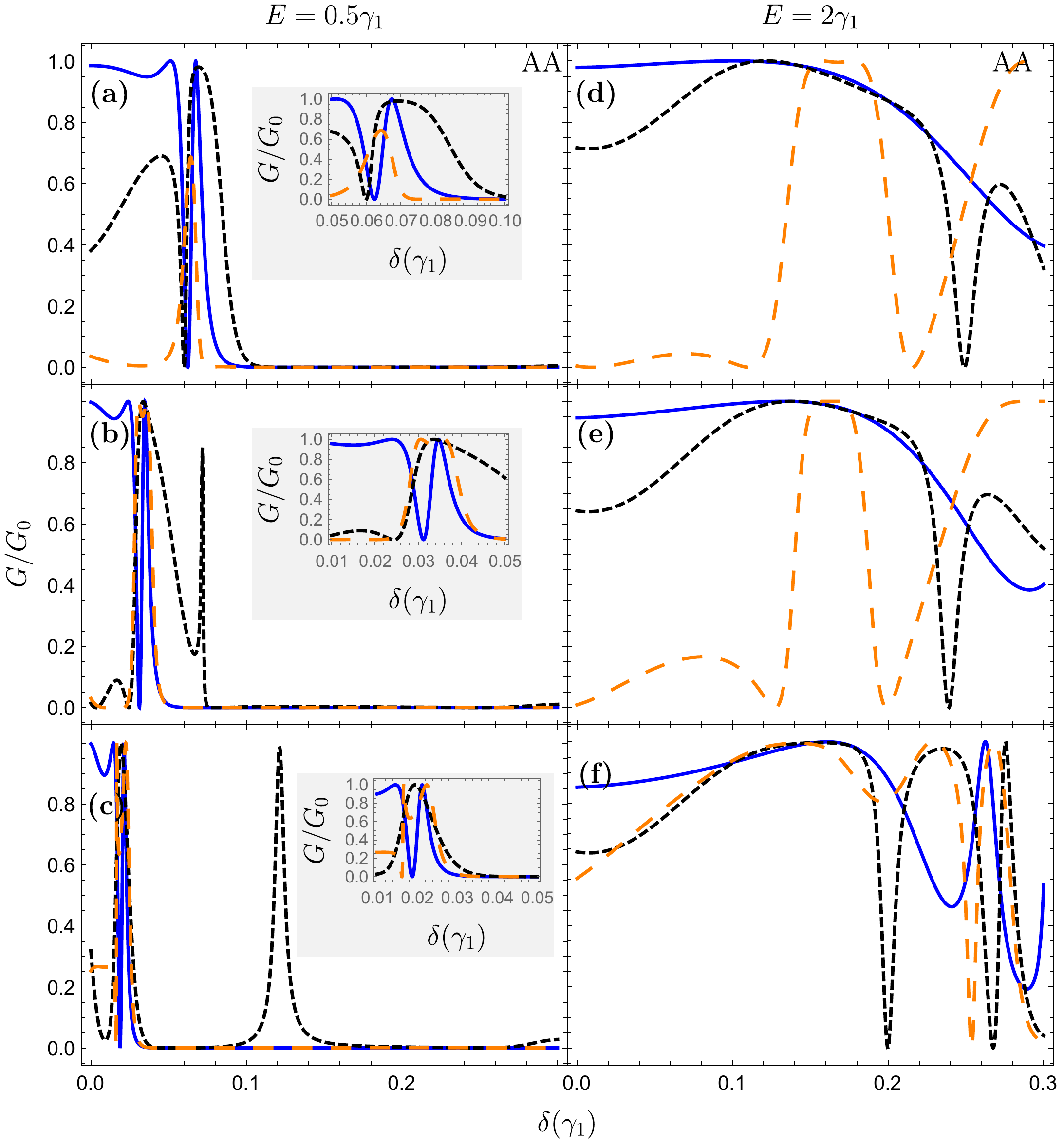}\includegraphics[width=0.45\linewidth]{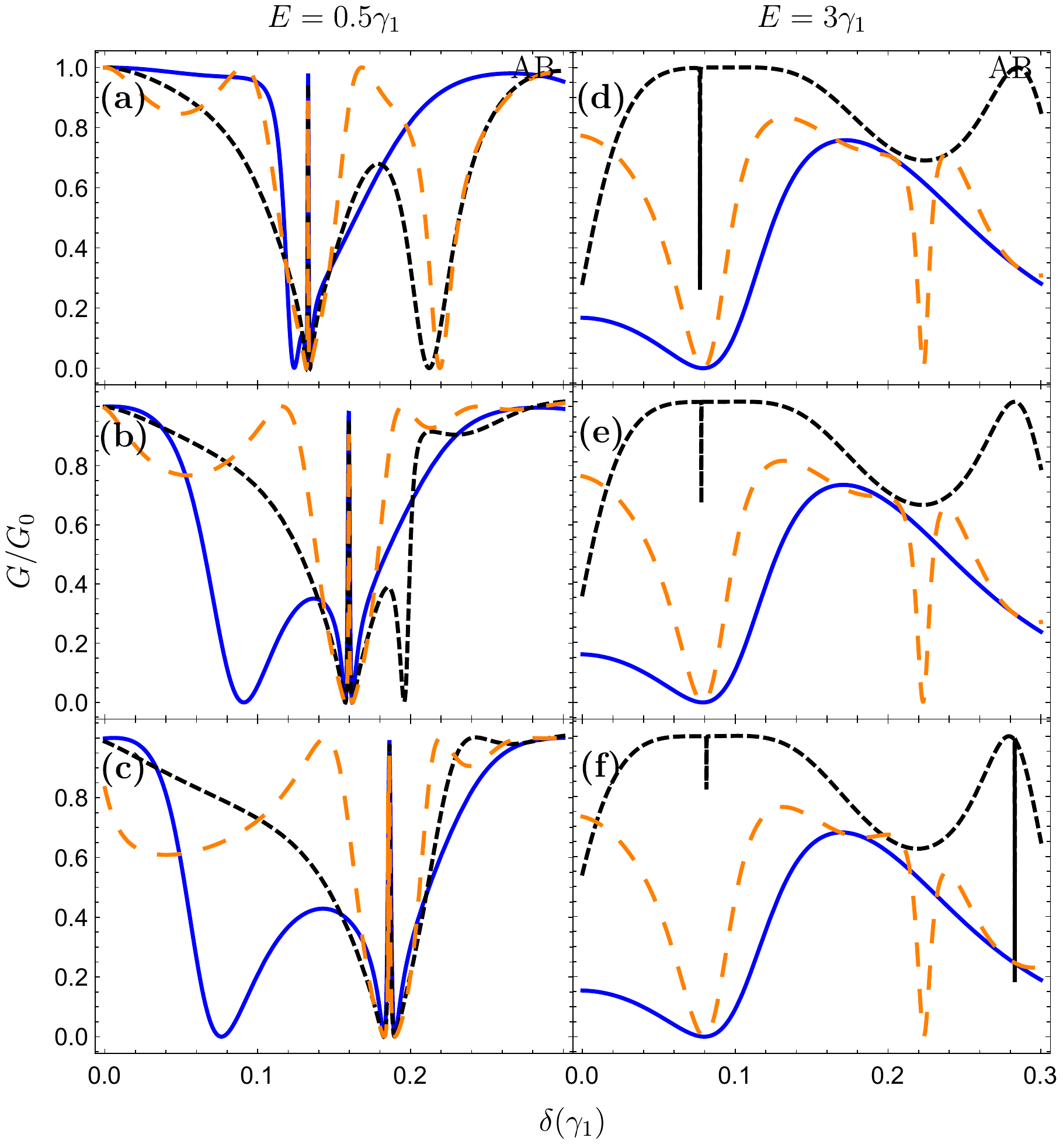}
	\caption{(Color online) Conductance for $\text{SLG}/\text{AA-BLG}+(\Delta, \delta)/\text{SLG}$, and  $\text{SLG}/\text{AB-BLG}+(\Delta, \delta)/\text{SLG}$ as a function of the bias $\delta$ for normal incidence, with different value of bilayer region length $d=10$ (blue solid line), $d=20$ (black dashed line), $d=30$ (orange dashed line), and parameters in left panel: (a) $(E, \Delta)=(0.5\gamma_1, 0)$, (b) $(E,  \Delta)=(0.5\gamma_1,  0.2\gamma_1)$, (c) $(E,  \Delta)=(0.5\gamma_1,  0.3\gamma_1)$, (d) $(E,\Delta)=(2\gamma_1,  0)$, (e) $(E,  \Delta)=(2\gamma_1,  0.2\gamma_1)$, (f) $(E,  \Delta)=(2\gamma_1,  0.3\gamma_1)$. In the right panel with parameters: (a) $(E, \Delta)=(0.5\gamma_1, 0)$, (b) $(E,  \Delta)=(0.5\gamma_1,  0.1\gamma_1)$, (c) $(E,  \Delta)=(0.5\gamma_1,  0.2\gamma_1)$, (d) $(E,\Delta)=(3\gamma_1,  0)$, (e) $(E,  \Delta)=(3\gamma_1,  0.1\gamma_1)$, (f) $(E,  \Delta)=(3\gamma_1,  0.2\gamma_1)$. }\label{conductance-as-function-of-delta-AA}
\end{figure*}
The conductance $G(E)$ as a function of the Fermi energy is shown in Figs. \ref{fig-conductance-energy2} (a,b) for two  values of the length $d=10$ (blue solid line), and $d=20$ (magenta dashed line)  with bias for $\text{SLG}/\text{AA-BLG}+(\Delta, \delta)/\text{SLG}$ and without bias for $\text{SLG}/\text{AB-BLG}+(\Delta, \delta)/\text{SLG}$ with $d=10$. A oscillating  dependence between the conductance $G$ and lengths larger than 10 is apparent. However, as depicted in the figures, the conductance  vanishes and shows anti-Klein tunneling when $d$ is smaller than $10$.
The density plot of transmission probability $T$  as a function of bilayer region length $d$ and Fermi energy $E$ through $\text{SLG}/\text{AA-BLG}+(\Delta, \delta)/\text{SLG}$, and $\text{SLG}/\text{AB-BLG}+(\Delta, \delta)/\text{SLG}$ junctions is shown in Fig. \ref{DP2}. The presence of a biased potential $\delta$ and a band gap $\Delta$ can significantly alter  $T$. Around $E = 0.25\gamma_1$, $T$ becomes asymmetric and strongly suppressed for low Fermi energy $E$. This is due to the fact that with an increase of the two parameters  $\delta$ and $\Delta$, the Dirac electron can be controlled within a finite region in both configurations \cite{wang2012tunable}.


In Fig. \ref{conductance-as-function-of-delta-AA}, we show the conductance in {$\text{SLG}/\text{AA-BLG}+(\Delta, \delta)/\text{SLG}$} as a function of the bias with (a,b,c) for $E=0.5\gamma_1$ and (d,e,f) for $E=2\gamma_1$. The results show anti-resonances for $\delta<0.1\gamma_1$ in Fig. \ref{conductance-as-function-of-delta-AA} (a,b). Furthermore, we see the conductance steadily decreasing for a large bias value independent of the bilayer region length $ d $.  
Selecting $\Delta=0.5\gamma_1$, we obtain zero-conductance with extra peaks appearing in the conductance profile with bias attributed to the transmitting channels in AA-BLG, see Fig. 
\ref{conductance-as-function-of-delta-AA} (c). 
Fig. \ref{conductance-as-function-of-delta-AA}  (d,e,f) depict clear anti-resonances.  The findings show that the conductance reflects opposite behavior as a function of the inter-layer bias and more specifically, the Fermi energy. {The conductance through $\text{SLG}/\text{AB-BLG}+(\Delta, \delta)/\text{SLG}$ case has distinct characteristics, it starts conducting with maxima for $E = 0.5\gamma_1$ and minima for $E = 3\gamma_1$, as shown in Fig. \ref{conductance-as-function-of-delta-AA} (right panel)}.

\section{Conclusion}
\label{Conclusion}
We have studied the conductance through single-layer graphene (SLG), and bilayer graphene (BLG) junctions, taking into account band gap $(\Delta)$ and bias ($\delta$) voltage terms. We started with gapped SLG with perfect AA/AB-BLG ($\text{SLG}+\Delta/\text{AA-BLG}/\text{SLG}+\Delta$), and ($\text{SLG}+\Delta/\text{AB-BLG}/\text{SLG}+\Delta$). We have shown that for Fermi energy larger than the interlayer hopping, the conductance as a function of the bilayer region of length reveals two different models of anti-resonances with the same period. As a function of the band gap, in AA-BLG stacking, the results show that the conductance has the same minima whatever the value of $d$, and for AB-BLG, $d$ remains relevant such that the system creates a global energy gap. 

In the second configuration, we considered pristine SLG and gapped-biased BLG ($\text{SLG}/\text{AA-BLG}+(\Delta, \delta)/\text{SLG}$), and ($\text{SLG}/\text{AB-BLG}+(\Delta, \delta)/\text{SLG}$). We found the appearance of peaks in the conductance profile with different periods and shapes, and also the presence of Klein tunneling with zero conductance in contrast to the first configuration. When $d$ is less than 10, the conductance $ G(E) $ vanishes and exhibits anti-Klein tunneling as a function of the Fermi energy $ E $. We have also evaluated the conductance as a function of the bias. For AA-BLG, the results show antiresonances and diminish for a large value of the bias independent of the bilayer region of length. On the other hand, in AB-BLG, conductance has distinct characteristics.


\bibliographystyle{unsrt}
\bibliography{mybib}
\appendix\label{Appendix}
\section{Gapped single layer graphene}
The eigenfunctions of the  gapped SLG Hamiltonian are given by
\begin{equation}
\psi(x,y)=\mathcal{G}_{\text{SLG}}\mathcal{M}(x)\mathcal{C}e^{ik_{y}y},
\end{equation}
where
\begin{equation}
\mathcal{G}_{\text{SLG}}=\begin{pmatrix}
\alpha^{-} & -\alpha^{+}  \\

1 & 1   \\
\end{pmatrix},\quad \mathcal{M}(x)=\begin{pmatrix}
e^{ik x}  \\
e^{-ikx}
\end{pmatrix},\quad \mathcal{C}=\begin{pmatrix}
a \\
b\\
\end{pmatrix},
\end{equation}
and 
\begin{equation}
k_{x}=\sqrt{E^{2}-\Delta^2-k_{y}^{2}},\quad \alpha^{\pm}=\frac{k_{x}\pm ik_{y}}{E-\Delta}.
\end{equation}
\section{AA-Wavefunction}
We demonstrate that the solution to Eq. (\ref{second-order-diff-equation}) is a plane wave formed by
\begin{equation}
\phi_{A_{1}}=a_{1}e^{ik^{+}x}+a_{2}e^{-ik^{+}x}+a_{3}e^{ik^{-}x}+a_{4}e^{-ik^{-}x},
\end{equation}
where $a_{n} $ are the coefficients of normalization, with $n = 1, 2, 3, 4$. The rest of the spinor components can be obtained by substituting this into equations (\ref{setofequation1}-\ref{setofequation4}). The system's wave function can be expressed in matrix form as
\begin{equation}
\psi(x,y)=\mathcal{G}_{\text{AA}}\mathcal{M}(x)\mathcal{C}e^{ik_{y}y},
\end{equation}
where
\begin{equation}
\mathcal{M}(x)=\begin{pmatrix}
e^{ik^{+}x} & 0 & 0& 0 \\
0 &e^{-ik^{+}x} & 0 & 0\\
0 &0  & e^{ik^{-}x} & 0\\
0 &0&  0& e^{-ik^{-}x} \\
\end{pmatrix},\quad  \mathcal{C}=\begin{pmatrix}
a1 \\
a2\\
a3\\
a4\\
\end{pmatrix},
\end{equation}
and 
\begin{equation}
	\mathcal{G}_{\text{AA}}=\begin{pmatrix}
	\chi^{+}_{+} & 	-\chi^{+}_{-} &\chi^{-}_{+}&\chi^{-}_{-} \\
1 & 1 & 1 & 1\\
	-\Lambda^{+}_{+} & \Lambda^{-}_{+} & -\Lambda^{+}_{-} & \Lambda^{-}_{-} \\
\xi^{+}_{+}	 & \xi^{+}_{-}&  \xi^{-}_{+}&\xi^{-}_{-} \\
	\end{pmatrix},
\end{equation}
where
\begin{align}
&\Lambda^{\pm}_{\alpha}=\frac{E^2\delta\pm E\lambda(\delta+\Delta)+\epsilon(\gamma_{1}^2\pm E\lambda)+\delta(\delta+\Delta)}{E\gamma_{1}(k_{x}^{\pm}+\alpha ik_y)},	\\
& \xi^{\pm}_{\alpha}=\frac{1}{\gamma_1}\left[E+\Delta-\delta-\alpha( k_{x}^{\pm}+ik_y)\chi_{\alpha}^{\pm}\right],\\
& \chi_{\alpha}^{\pm}=\chi_{\alpha}^{\pm} =\frac{1}{E-\Delta-\delta} \left[k_{x}^{\pm}-\alpha ik_y-\gamma_1\Lambda^{\pm}_{\alpha}\right],
\end{align}
with $\alpha=\pm$.
\section{AB-Wavefunction}
The wave function of the system can be written in matrix form as
\begin{equation}
\psi(x,y)=\mathcal{G}_{\text{AB}}\mathcal{M}(x)\mathcal{C}e^{ik_{y}y},
\end{equation}
where the four-component vector $C$ represents the different coefficients, with the matrix $\mathcal{G}_{\text{AB}}$ given by
\begin{equation}
\mathcal{G}_{\text{AB}}=\left(\begin{array}{cccc}
\chi^{+}_{+} & \chi^{+}_{-} & \chi^{-}_{+} & \chi^{-}_{-} \\
1 & 1 & 1 & 1 \\
-\Lambda^{+}_{+} &  \Lambda^{-}_{+} & -\Lambda^{+}_{-} & \Lambda^{-}_{-} \\
\rho^{+}_{+} & \rho^{+}_{-} & \rho^{-}_{+} & \rho^{-}_{-}
\end{array}\right),
\end{equation}
and we have set
\begin{align}
&\Lambda^{s}_{\pm}=\frac{(k^{s}\pm ik_{y})}{\varepsilon-\delta-\Delta},\\
&\rho^{\pm}=\frac{(\varepsilon-\delta-\Delta)(\varepsilon-\delta+\Delta)-(k_y)^2-(k^{\pm})^2}{\Gamma_{1}(\varepsilon-\delta-\Delta)},\\ &\xi^{\pm}_{\alpha}=\frac{\alpha k^{\pm}-ik_y}{\varepsilon+\delta+\Delta}.
\end{align}
\section{Wave-function matching}\label{Appendix-D}
In the configuration adopted, the wave-function is continuous in the bottom layer \cite{neto2009electronic}
\begin{equation}
\begin{aligned}
& \begin{cases}\phi_{\text{A}_1}(x=0)=\phi_{\text{B}_1}(x=0), \\
\phi_{\text{A}_2}(x=0)=\phi_{\text{B}_2}(x=0)=0,\\
\phi_{\text{A}_1}(x=d)=\phi_{\text{B}_1}(x=d), \\
\phi_{\text{A}_2}(x=d)=\phi_{\text{B}_2}(x=d)=0. \\
\end{cases}
\end{aligned}
\end{equation}
\end{document}